\documentclass[a4paper,onecolumn,11pt,accepted=2023-09-05]{quantumarticle}

\pdfoutput=1

\usepackage[utf8]{inputenc}
\usepackage[english]{babel}
\usepackage[T1]{fontenc}
\usepackage{dirtytalk} %simple quotation marks with \say{}
\usepackage{amsmath}

\usepackage{hyperref} % For refs and labels
\usepackage{cleveref}
\usepackage{physics} % For \ket etc 
\usepackage[normalem]{ulem} % For strikethrough
\usepackage{tikz} % For tikz pictures
\usepackage{lipsum} % For abstract place holder
\usepackage{wrapfig}%for figures floating in text
\usepackage{amssymb} % For \leqslant
\usepackage{framed} % For frames around pictures (including their captions)
\usepackage{tabularx} % For the protocol environment to work properly

\newtheorem{theorem}{Theorem}
\newtheorem{definition}{Definition}
\newtheorem{lemma}{Lemma}
\newtheorem{corollary}{Corollary}

\newcommand{\GHZ}[1]{\ensuremath{\mathrm{GHZ}_{#1}}} % GHZ states
\newcommand{\Lin}[1]{\ensuremath{\ket{\mathrm{L}_{#1}}}} % Linear cluster states

\newcommand{\node}[1]{\ensuremath{N_{#1}}} % Node definition

\newcommand{\Alice}[0]{$\node{a}$} % Alice
\newcommand{\Bob}[0]{$\node{b}$} % Bob
\newcommand{\Charlie}[0]{$\node{c}$} % Charlie

\newcommand{\qubt}[1]{\ensuremath{\tau_{#1}}} % Top qubit for a node in Bell pair with node on the right
\newcommand{\qubb}[1]{\ensuremath{\omega_{#1}}} % Bottom qubit for a node in Bell pair with node on the left
\newcommand{\qub}[1]{\ensuremath{#1}} % Remaining qubit after the state preparation, part of a lincluster state
\newcommand{\qubr}[1]{\ensuremath{\tilde{#1}}} % Rest-qubit, there for Alice and Charlie who don't measure at state prep

\newcommand{\spmo}[1]{\ensuremath{o_{#1}}} % State preparation measurement outcome
\newcommand{\gemo}[1]{\ensuremath{m_{#1}}} % GHZ extraction measurement outcome
\newcommand{\gemb}[1]{\ensuremath{\beta_{#1}}} % GHZ extraction measurement basis bit

\newcommand{\net}[0]{\ensuremath{\mathcal{N}}} % Full network
\newcommand{\partic}{\ensuremath{\mathcal{P}}} % Participants
\newcommand{\nonpart}[0]{\ensuremath{\bar{\mathcal{P}}}} % Non-participants

\newcommand{\ie}{{i.e.~}} % Good-looking i.e.
\newcommand{\eg}{{e.g.~}} % Good-looking e.g.

\newcommand{\fourgroup}[0]{\ensuremath{g_{ab}}} % Symbol for the number of four-groups in \delta_{ab}
\newcommand{\modfourval}[0]{\ensuremath{p_{ab}}} % Symbol for the mod-four value of \delta_{ab}

\newcommand{\hmin}{H_{\mathrm{min}}}
\newcommand{\hmax}{H_{\mathrm{max}}}

\newcounter{protocol}
\newenvironment{protocol}[1]
 {\par\addvspace{\topsep} % Set the distance between the text above and the first hline of the protocol environment
 \noindent
 \tabularx{\linewidth}{@{} X @{}} % You need to load the tabularx package for this to work
 \refstepcounter{protocol}\textbf{Protocol \theprotocol: #1} \\
 }
 {
 \endtabularx
 }

\def\defeq{\mathrel{\mathop:}=} %definitions

\definecolor{pink}{RGB}{237,16,118} %Charlie
\definecolor{applegreen}{rgb}{0.55, 0.71, 0.0} %Alice
\definecolor{celestialblue}{RGB}{62,146,204} %Bob
\definecolor{lilanew}{RGB}{152,41,221}
\definecolor{colormult00}{RGB}{86, 180, 233}%light blue
\definecolor{colormult01}{RGB}{230, 159, 0}%orange
\definecolor{colormult02}{RGB}{204, 121, 167}%lila
\definecolor{colormult03}{RGB}{240, 228, 66}%yellow
\definecolor{colormult04}{RGB}{0, 158, 115}%green
\definecolor{colormult05}{RGB}{213, 94, 0}%kindo of orange-red
\definecolor{colormult06}{RGB}{0, 114, 178}%dark blue
\definecolor{colormult07}{RGB}{0,0,0} %just black
\definecolor{violet}{HTML}{53257F} %Quantum violet
\definecolor{green}{HTML}{257a7f}%{507F25} %Quantum green
\definecolor{brown}{HTML}{852e29}%{7F3B25} %Quantum brown

\begin{document}
% \title{Anonymous conference key agreement \\
% %\par %$\hspace{0.3\textwidth}$
% in linear quantum networks}

\title{Anonymous conference key agreement in linear quantum networks}

\author{Jarn de Jong}
\email{dejong@tu-berlin.com}
\affiliation{Electrical Engineering and Computer Science, Technische Universit{\"a}t Berlin, 10587 Berlin, Germany}
\orcid{0000-0001-9662-9337}
\author{Frederik Hahn}
\affiliation{Electrical Engineering and Computer Science, Technische Universit{\"a}t Berlin, 10587 Berlin, Germany}
\affiliation{Dahlem Center for Complex Quantum Systems, Freie Universität Berlin, 14195 Berlin, Germany}
\orcid{0000-0002-9349-4075}
\author{Jens Eisert}
\affiliation{Dahlem Center for Complex Quantum Systems, Freie Universität Berlin, 14195 Berlin, Germany}
\orcid{0000-0003-3033-1292}
\author{Nathan Walk}
\affiliation{Dahlem Center for Complex Quantum Systems, Freie Universität Berlin, 14195 Berlin, Germany}
\orcid{0000-0003-1204-6009}
\author{Anna Pappa}
\affiliation{Electrical Engineering and Computer Science, Technische Universit{\"a}t Berlin, 10587 Berlin, Germany}
\orcid{0000-0002-4662-149X}

\maketitle

%%%%%%%%%%%%%%%%%%%%%%%%%%%%%%%%%%%%%%%%%%%%%%% ABSTRACT %%%%%%%%%%%%%%%%%%%%%%%%%%%%%%%%%%%%%%%%%%%%%%%%%%%%%%%%%%%
\begin{abstract}
 \noindent
 Sharing multi-partite quantum entanglement between parties allows for
 diverse secure communication tasks to be performed. Among them, conference key agreement (CKA) -- an extension of key distribution to multiple parties -- has received much attention recently. Interestingly, CKA can also be performed in a way that protects the identities of the participating parties, therefore providing \emph{anonymity}. 
 In this work, we propose an anonymous CKA protocol for three parties
 that is implemented in a highly practical network setting. Specifically, a line of quantum nodes is used to build a linear cluster state among all nodes, which is then used to anonymously establish a secret key between any three of them. The nodes need only share maximally entangled pairs with their neighbours, therefore avoiding the necessity of a central server sharing entangled states. This linear chain setup makes our protocol an excellent candidate for implementation in future quantum networks. We explicitly prove that our protocol protects the identities of the participants 
  from one another
 and perform an analysis of the key rate in the finite regime,
 contributing to the quest of identifying feasible quantum communication tasks for network architectures beyond point-to-point.
\end{abstract}
\vspace{-1.3cm}

%%%%%%%%%%%%%%%%%%%%%%%%%%%%%%%%%%%%%%%%%%%%%%% INTRODUCTION %%%%%%%%%%%%%%%%%%%%%%%%%%%%%%%%%%%%%%%%%%%%%%%%%%%%%%%%%%%
\section{Introduction}
The goal of \emph{conference key agreement} (CKA) protocols is to establish a shared key between multiple parties that do not use trusted means of communication. 
%This primitive can be seen as an extension of key agreement to include more than two parties. 
CKA has been explored in the quantum domain, 
%\cite{murta2020quantum} 
with the aim of developing new schemes for cryptography %\cite{Advances,Roadmap} 
and communication %\cite{Principles} 
beyond bi-partite key distribution \cite{murta2020quantum, Pirandola:20,Roadmap, khatri2020principles}. 
Various quantum states have been proposed to achieve CKA, including both
\textit{discrete-variable} %(DV) 
\cite{epping_multi-partite_2017, grasselli_finite-key_2018, grasselli2019conference}
and 
\textit{continuous-variable} %(CV)  
\cite{wu2016continuous, ottaviani2019modular, zhang2018multipartite} states. 
% For the DV approach, particularly useful are \GHZ{} states \cite{greenberger1989bell} which can be employed as a resource for CKA protocols due to their obvious correlation when measured in the computational basis \cite{murta2020quantum, epping_multi-partite_2017, grasselli_finite-key_2018}. Other states that have been used are so-called $W$ states, albeit only allowing to create a conference key probabilistically \cite{grasselli2019conference}. Both of these resources are symmetric in what concerns the correlations between the participants of the protocol, and it is assumed that they are created and shared by a central node. 
In this work, we focus on the discrete-variable case.
Particularly suitable for CKA protocols are quantum resource states that exhibit symmetric measurement outcome correlations for all participants of the respective communication protocol.
Such states include \GHZ{} states \cite{greenberger1989bell}, which are ideal CKA resources due to their obvious correlation when measured in the computational basis \cite{murta2020quantum, epping_multi-partite_2017, grasselli_finite-key_2018}. 
Other useful symmetric resources are $W$ states \cite{WstatePhysRevA.62.062314}, although they only allow for the probabilistic generation of a conference key \cite{grasselli2019conference}. 

In recent years, the need for \textit{privacy and anonymity} has led researchers to develop anonymous versions of protocols that implement important cryptographic primitives \cite{huang2022experimental, Yang2021Towards, unnikrishnan_anonymity_2019}. In the case of CKA, the goal would be to establish a key between a number of participants, while keeping their identities hidden from the other non-participating parties, and in some instances even from each other. 
Using shared GHZ states, anonymous protocols have been proposed \cite{hahn2020anonymous,grasselliSecureAnonymousConferencing2022} and implemented \cite{thalacker_anonymous_2021} that also provide an advantage in keyrate compared to sharing bi-partite entanglement \cite{grasselliSecureAnonymousConferencing2022}.
However, GHZ states are highly loss-prone and cannot be easily exchanged over long distances. 
Moreover, most previous approaches using these states require a central server to distribute the state.

In this work, we address and answer the question of how to achieve anonymous CKA within a minimal and in several ways experimentally feasible nearest neighbour architecture of discrete variable quantum states.
Specifically, we assume a linear chain of quantum nodes\footnote{Note that “linear chain” setups often refer to quantum repeater setups involving quantum memories. Our protocol however does not require quantum memories. The feasibility of implementing this protocol in the short term, and the relevance of the photonics experiments discussed in Section~\ref{sec:discussion}, are due to the fact that the setup is a linear chain of source and measurement stations without quantum memories, rather than a traditional linear repeater chain.}, and study how to anonymously share a secret key between three of them. 
These three participants (also referred to as Alice, Bob and Charlie) use shared bi-partite entanglement with their neighbours to establish a linear cluster state
\cite{Briegel-PRL-2001,Oneway,Graphs}
that `connects' the three of them and all the nodes in between. This linear cluster state is further used to anonymously establish a three-partite maximally entangled state between Alice, Bob and Charlie, which is used to run a conference key agreement protocol and share a key with provable security. Our new protocol is one of the few cryptographic demonstrations where non-maximally entangled states are used for practical tasks. It highlights the flexibility that multi-partite entanglement provides for cryptographic tasks involving more than two parties and complements known results supporting this approach \cite{hahn2020anonymous, grasselliSecureAnonymousConferencing2022, christandl2005quantum, walk2021sharing, markham2008graph}.

This work is structured as follows. In Subsection \ref{subsec:preliminaries} we provide the necessary preliminaries and definitions to subsequently introduce the protocol in Section \ref{sec:protocol}. Section \ref{sec:analysis} contains an analysis and explanation of the protocol and its performance, specifically how anonymity and security are achieved, as well as how the (finite) key rate is calculated. Section \ref{sec:discussion} contains a general discussion. Some details of the protocol, as well as the proofs of security and anonymity have been deferred to Appendices \ref{append:corrections}, \ref{append:security} and \ref{append:anonymity}, respectively.

\subsection{Preliminaries and definitions}\label{subsec:preliminaries}
\begin{figure}[ht!]
\begin{framed}
\begin{centering} 
	\includegraphics[trim=
 6 %left
 6 %bottom 
 14 %right 
 6 %top
 , clip,width=\textwidth]{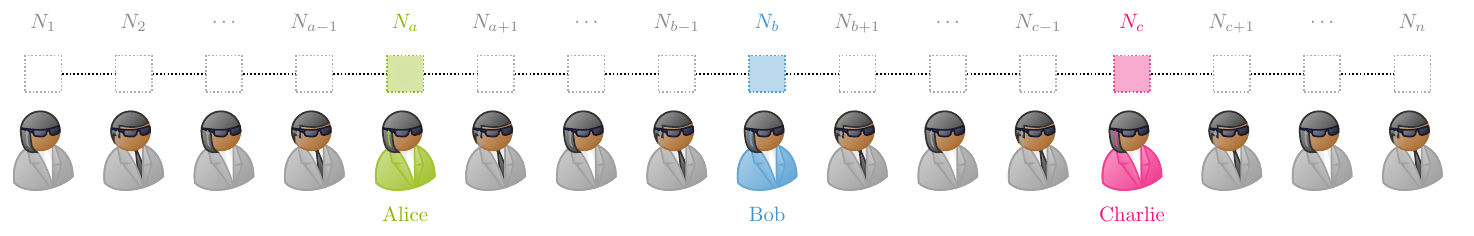}
	\caption{Participants $\partic = \{\textcolor{applegreen}{\node{a}},\textcolor{celestialblue}{\node{b}},\textcolor{pink}{\node{c}}\}$ 
% 	(\ie Alice, Bob and Charlie) 
	and non-participants $\nonpart$ are connected by a linear network. The position of Alice, Bob and Charlie are not known by any other node in the network.}
	\label{fig:networklayout}
\end{centering}
\end{framed}
\end{figure}

Throughout this work, we denote with the ordered set
$\net = \{\node{1},\node{2},\ldots,\node{n-1},\node{n}\}$
the collection of all nodes in a linear network (Fig.~\ref{fig:networklayout}). We fix a set of three nodes $\partic = \{\node{a},\node{b},\node{c}\} \subset \net$ such that $ a < b < c$ and call them \textit{Alice}, \textit{Bob} and \textit{Charlie}, respectively. 
The goal of these three parties is to establish a secret key between them, without revealing their identities to everyone else. We will refer to them as the \textit{participants} and to all other nodes $\nonpart \defeq \net \setminus \partic$ as the \textit{non-participants} of our anonymous conference key agreement protocol.

\vspace{-0.2in}
Alice, Bob and Charlie know each other's position within the network and can identify the communication from each \node{i} with its index $i$.
Moreover, they have access to a pre-shared secret key; note that our scheme is a key-expansion scheme, like all known quantum key distribution and CKA protocols.
We assume that the participants are cooperating: neither do they reveal each other's identity nor do they reveal the created secret key. However, we allow the non-participants to be dishonest and to actively deviate from the protocol -- as long as they do not collude with each other. 
This means that non-participants can perform arbitrary maps and measurements in deviating bases or disclose false measurement outcomes, but they cannot, jointly with other non-participants, perform a coordinated attack. 

\begin{figure}[ht!]
\begin{framed}
\begin{centering}
 \includegraphics[trim=
 4 %left
 7 %bottom 
 27 %right %too much white space here
 2 %top
 , clip, width=\textwidth]{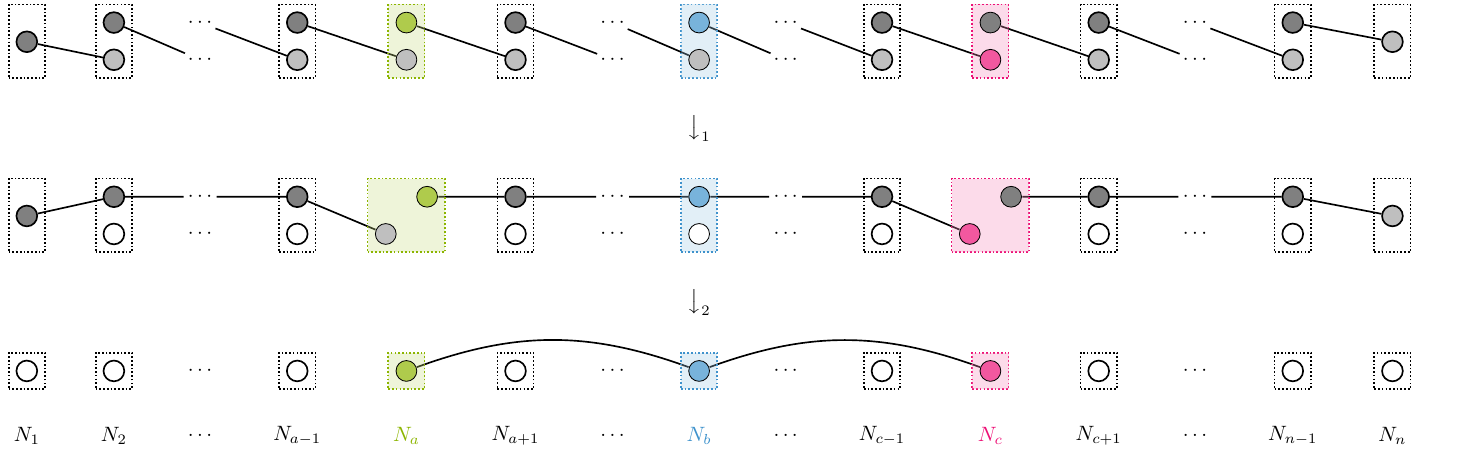}
	\caption{\textbf{Top:} Bell pairs shared between nodes $\node{i}$ and $\node{i+1}$. Qubit $\qubt{i}$ (\textcolor{gray}{dark gray}) of \node{i} and $\qubb{i+1}$ (\textcolor{gray!50}{light gray}) of \node{i+1} are entangled. 
	The three qubits to be part of the \GHZ{} state are colored \textcolor{applegreen}{green}, % (Alice),
	\textcolor{celestialblue}{blue} % (Bob) 
	and \textcolor{pink}{pink}. % (Charlie).
	\textbf{Middle:} 
	In Protocol~\ref{prot:statepreparation}, the Bell resources are used to create three linear cluster states via \textit{Bell state projection}. Alice and Charlie do not perform the projection.
	\textbf{Bottom:} In Protocol~\ref{prot:ghzextraction} the network states $\ket{\net}$ are transformed into $\ket{\GHZ{3}}$ states between \textcolor{applegreen}{\Alice}, \textcolor{celestialblue}{\Bob}~and \textcolor{pink}{\Charlie}.
	}
	\label{fig:statesduringprotocol}
\end{centering}
\end{framed}
\end{figure}

Regarding initial resources, we assume that all nodes share one copy of an entangled
(potentially noisy)
\textit{Bell pair} -- \ie 
% the state 
$(\ket{0,0}+\ket{1,1})/\sqrt{2}$ -- with each of their neighbours: node $\node{i}$ therefore holds two qubits with labels $\qubt{i}$ and $\qubb{i}$. Qubit $\qubt{i}$ is entangled with qubit $\qubb{i+1}$ from node $\node{i+1}$ while qubit $\qubb{i}$ is entangled with qubit $\qubt{i-1}$ from $\node{i-1}$. Since nodes $\node{1}$ and $\node{n}$ both have a single neighbour, they only have a qubit $\qubt{1}$ and $\qubb{n}$, respectively.
These Bell pairs are used to create three \textit{linear cluster states} between the nodes. For a set of qubits $\{1,2,\dots, n\}$, we define the $n$-qubit linear cluster state vector
\begin{equation}
 \Lin{1,\dots, n} \defeq \prod_{i=1}^{n-1}CZ_{i,i+1}\ket{+}^{\otimes n},
\end{equation}
where $\ket{+}^{\otimes n} \defeq \bigotimes_{i=1}^{n}\ket{+}_{i}$ and $CZ_{i,i+1}$ is the controlled-$Z$ quantum gate that is applied between neighbouring qubits labelled $i$ and $i+1$.
Finally, from the linear cluster state, a three-partite $\GHZ{}$ state with state vector
\begin{equation}
\ket{\GHZ{3}} := \frac{1}{\sqrt{2}}\left(
\ket{0,0,0} +
\ket{1,1,1}
\right)
\end{equation}
is extracted in an anonymous fashion. Fig.~\ref{fig:statesduringprotocol} depicts a visualization of our protocol.
For clarity, throughout this work, we denote a unitary Pauli operation with a capital letter, where a superscript indicates a real power (\eg $Z^{b}$ is a unitary $Z$ operation for $b = 1$ and $I$ for $b = 0)$. To indicate measurement bases, we use the $\sigma$-notation, where a superscript indicates the qubit (\ie $\sigma_{x}^{i}$ refers to a measurement of qubit $i$ in the Hadamard basis).

%%%%%%%%%%%%%%%%%%%%%%%%%%%%%%%%%%%%%%%%%%%%% SECTION PROTOCOL %%%%%%%%%%%%%%%%%%%%%%%%%%%%%%%%%%%%%%%%%%%%%%%%%%%%
\section{Protocol} \label{sec:protocol}
Our protocol is divided into three parts -- the preparation of the required multi-partite states from the Bell pairs (\ref{subsec:preparation}), the anonymous extraction of the \GHZ{3} states from the multi-partite states (\ref{subsec:extraction}), and the subsequent key generation with post-processing (\ref{subsec:measure_post}).

\vspace{-0.2cm}
% \vspace{-0.9cm}

\subsection{Preparation of linear cluster states}\label{subsec:preparation}
From the Bell resources, we create the \textit{network state} vector 
$
 \ket{\net} \defeq \Lin{\mathrm{l}} \otimes \Lin{\mathrm{m}} \otimes \Lin{\mathrm{r}},
$
where $\Lin{\mathrm{l}}$, $\Lin{\mathrm{m}}$ and $\Lin{\mathrm{r}}$ -- with $\mathrm{l, m, r}$ denoting left, middle and right -- are linear cluster states on the qubits labelled $\{\qubt{1},\qubt{2},\dots, \qubt{a-1},\qubb{a}\}$, $\{\qubt{a},\qubt{a+1},\dots, \qubt{c-1},\qubb{c}\}$ and $\{\qubt{c},\qubt{c+1},\dots, \qubt{n-1},\qubb{n}\}$, respectively (Fig.~\ref{fig:statesduringprotocol} top and middle), by performing Protocol~\ref{prot:statepreparation} for \texttt{State preparation}.

Protocol~\ref{prot:statepreparation} is divided into three steps. Some nodes execute only a subset of the steps and a small selection of nodes execute a variant of Step $2a$, listed as Step $2b$. These considerations are reflected in Tab.~\ref{tab:statepreparation}, which indicates which nodes perform which steps.

%%%%%%%%%%%%%%%%%%%%%%%%%%%%%% PROTOCOL 1 - STATE PREPARATION %%%%%%%%%%%%%%%%%%%%%%%%%%%%%%
% \vspace{-0.2cm}
\begin{framed}
\noindent\begin{minipage}[htpb]{0.56\textwidth}
\vspace{-0.33cm}
\begin{protocol}{\label{prot:statepreparation} \texttt{State preparation}}
\hrule
\textit{Input.} 
Bell pairs between nodes $\node{i}$ and $\node{i+1}$.
\newline \textit{Goal.} Prepare the network state vector $\ket{\net}$.

\vspace{\baselineskip}
\noindent All $\node{i}$ perform the following steps consecutively:
\begin{enumerate} \itemsep0.1em
 \item[1.] 
 \begin{itemize} \itemsep0em \label{subprotocol1:stepreceiveandcorrect} % Shorter distances between items
 \item[] Receive $\spmo{i-1}$.
 \item[] If $\spmo{i-1} = 1$ apply $Z$ on $\qubb{i}$.
 \end{itemize}
 \item[2a.] \begin{itemize} \itemsep0em 
 \item[] Perform $CZ$ between \qubt{i} and \qubb{i}.
 \item[] Measure $\sigma_{x}^{\qubt{i}}$ and record measurement outcome bit as \spmo{i}.
 \end{itemize}
 \item[2b.] \begin{itemize} \itemsep0em 
 \item[] Draw uniformly random bit \spmo{i}. 
 \item[] If $\spmo{i} = 1$ apply $Z$ on $\qubt{i}$. 
 \item[] Apply $H$ on $\qubt{i}$.
 \end{itemize}
 \item[3.] \begin{itemize} \itemsep0em 
 \item[] Send \spmo{i} to \node{i+1}. \end{itemize}
\end{enumerate}
\end{protocol}
\vspace{-0.7cm}
\end{minipage}
\hspace{0.01\textwidth}
\begin{minipage}[t!]{0.4\textwidth}
 \captionof{table}{Protocol~\ref{prot:statepreparation} Overview \\
 \newline
 Not all steps of Protocol~\ref{prot:statepreparation} are performed by everyone. This table indicates which steps are performed by whom.
 Note that since \node{1} does not perform 1.~and \node{n} does not perform 3.~there is neither $o_{0}$ nor \node{n+1}.}
 \includegraphics[width=\textwidth]{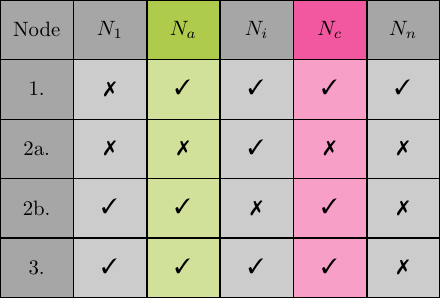}
 \label{tab:statepreparation}
\end{minipage}%
\hspace{0.03\textwidth}
\end{framed}
% \vspace{-0.2cm}

%%%%%%%%%%%%%%%%%%%%%%%%%%%%% END OF PROTOCOL 1 %%%%%%%%%%%%%%%%%%%%%%%
After discarding all measured qubits, each node but Alice and Charlie holds only one qubit. Therefore, we can rename $\qubt{i} \rightarrow \qub{i}$ for all $\node{i} \in \net \setminus \{\node{a},\node{c},\node{n}\}$ and $\qubb{n} \rightarrow \qub{n}$. For Alice and Charlie we rename \qubt{a} and \qubb{c} to \qub{a} and \qub{c} as part of \Lin{\mathrm{m}} and \qubb{a} and \qubt{c} to \qubr{a} and \qubr{b} as part of \Lin{\mathrm{l}} and \Lin{\mathrm{r}}, respectively. From the network state we will now show how to anonymously extract a \GHZ{3} state for Alice, Bob and Charlie.
Protocol \ref{prot:statepreparation} and the above discussion is phrased under the assumption that neither Alice nor Charlie are at the `ends' of \net. If indeed $\node{a} = \node{1}$, Alice performs the steps of the column marked \node{1} in Tab.~\ref{tab:statepreparation}, and similarly for Charlie if $\node{c} = \node{n}$. Note that Alice and/or Charlie then also hold just one qubit.

As a small example, consider a network with only five nodes, where $\partic{} = \{\Alice,~\Bob,~\Charlie\} = \{\node{2},\node{4},\node{5}\}$ and thus $\nonpart{} = \{\node{1}, \node{3}\}$. 
Note that $\node{n} =~\Charlie{}$.
All nodes except the first apply $Z$ to their qubit $\qubb{i}$ based on the outcome bit \spmo{i-1} they received.
Any node other than \Alice, \Charlie{} or the first or last node performs step $2a.$: they apply a $CZ$ gate and then measure $\sigma_{x}^{\qubt{i}}$, sending the outcome $\spmo{i}$ to the next node. 
To disguise that they are not performing this step, \node{1} and \Alice{} apply $Z$ to their \qubt{i} based on randomly drawn bit \spmo{i} and send it to the next node.
Moreover, they apply $H$ to \qubt{i}. Since \Charlie{} is the last node, they only perform step $1$.

% \vspace{-.4cm}
%%%%%%%%%%%%%%%%%%%%%%%%%%%%%%%%%%%%%%%% Subsection 2 GHZ extraction %%%%%%%%%%%%%%%%%%%%%%%%%%%%%%%%%%%%%%%%%
\subsection{Anonymous extraction of \texorpdfstring{\GHZ{}}{GHZ} states}
\label{subsec:extraction}
% \vspace{-.2cm}

After the $n$ network nodes have created $L$ network states %$\ket{\net}$ 
as presented in Protocol~\ref{prot:statepreparation}, each of these states is used to anonymously establish a $\ket{\mathrm{GHZ}_3}$ state between Alice, Bob and Charlie via Protocol~\ref{prot:ghzextraction} for \texttt{GHZ extraction}, as visualized in Fig.~\ref{fig:statesduringprotocol}.

%%%%%%%%%%%%%%%%%%%%%%%%%%%%%%%% PROTOCOL 2 GHZ EXTRACTION %%%%%%%%%%%%%%%%%%%%%%%%%
% \vspace{-.2cm}
\begin{framed}
\noindent\begin{minipage}[htpb]{0.627\textwidth}
\vspace{-0.33cm}
\begin{protocol}
{\label{prot:ghzextraction} \texttt{$\mathbf{GHZ}$ extraction}}
\hrule
\textit{Input.} Network state vector  $\ket{\net}$ from Protocol~\ref{prot:statepreparation}. \textit{Configuration} corrections $\{C^{i}\}_{i \in \partic}$ calculated by \node{i}, as explained in App.~\ref{append:corrections}.
\newline \textit{Goal.} Anonymous $\ket{\GHZ{3}}$  for $\node{a},\node{b},\node{c}$.

\vspace{\baselineskip}
\noindent All $\node{i}$ perform the following steps consecutively:
\begin{enumerate} \itemsep0.1em
 \item[1.] \begin{itemize} \itemsep0em
 \item[] Receive bit $\beta_{i-1}$ and compute $\beta_{i} = \beta_{i-1} \oplus 1$.
    \end{itemize}
 \item[2a.]
 \begin{itemize} \itemsep0em
    \item[] Measure $\sigma_{x}^{i}$ or $\sigma_{y}^{i}$ if $\gemb{i}$ is $0$ or $1$, respectively.
    \item[] Record the measurement outcome bit $\gemo{i}$.
 \end{itemize}
 \item[2b.] 
 \begin{itemize} \itemsep0em
 \item[] Draw a uniformly random bit $m_{i}$.
 \item[] If $i \in \partic{}$: apply  $C^{i}$.
 \end{itemize}
 \item[3.] \begin{itemize} \itemsep0em
 \item[] Communicate $\gemb{i}$ to node \node{i+1}.
    \end{itemize}
\end{enumerate}
\end{protocol}
% \vspace{-0.7cm}
\end{minipage}
\hspace{0.01\textwidth}
\begin{minipage}[t!]{0.333\textwidth}
 \captionof{table}{Protocol~\ref{prot:ghzextraction} Overview \\
 \newline
 \node{1} draws a uniformly random bit \gemb{1}; \node{n} does not perform Step $3$.}
\includegraphics[width=\textwidth]{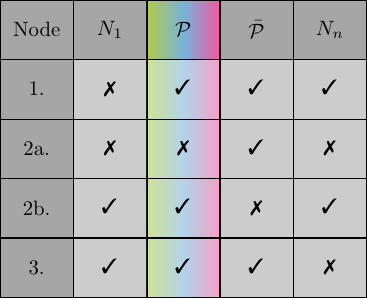}
 \label{tab:ghzextraction}
 \vspace{-0.4cm}
\end{minipage}
\hspace{0.03\textwidth}
\end{framed}
%%%%%%%%%%%%%%%%%%%%%%%%%%%% END OF PROTOCOL 2 %%%%%%%%%%%%%%%%%%%%%%%%%%%%%%%%%%

Protocol~\ref{prot:ghzextraction} is divided into three steps as well; some nodes only execute a subset of the steps and for Step $2$ there are again two different options; this is reflected in Tab.~\ref{tab:ghzextraction}. After a given amount of time for everyone to measure their qubit, all nodes broadcast their measurement outcome $\gemo{i}$. The participants $\partic$ then perform local unitary corrections on their own qubits based on the number of nodes between Alice, Bob and Charlie as well as the collection of measurement outcomes $\{\gemo{i}\}_{i \in \nonpart}$, resulting in $L$ \GHZ{3} states shared between them. These corrections can be found in App.~\ref{append:corrections} and consist of Clifford operations only. Importantly, the corrections invoked by the measurement outcomes can be accounted for in post-processing, so that all actions of the protocol can be carried out simultaneously; this ensures no quantum memories are necessary.
The final steps that enable secure anonymous conference key agreement are explained in the next section. 

Similarly to the previous one, Protocol \ref{prot:ghzextraction} is phrased under the assumption that neither Alice nor Charlie are at the `ends' of \net. If indeed $\node{a} = \node{1}$, Alice performs the steps of the column marked \node{1} in Tab.~\ref{tab:statepreparation}, and similarly for Charlie if $\node{c} = \node{n}$. Note that Alice and/or Charlie then also hold just one qubit.

Consider again the example where $\partic{} = \{\Alice,~\Bob,~\Charlie\} = \{\node{2},\node{4},\node{5}\}$, $\nonpart{} = \{\node{1}, \node{3}\}$ and 
$\node{n} =~\Charlie{}$. 
To start Protocol~\ref{prot:ghzextraction}, \node{1} draws uniformly random bits \gemb{1} and \gemo{1}. They send \gemb{1} to \node{2}.
Here, \node{2} is a participant and therefore does not measure their qubit after calculating $\beta_{2} = \beta_{1} \oplus 1$. Instead they apply $C^{2}$, draw the bit \gemo{2} uniformly and send \gemb{2} to \node{3}.
The following node \node{3} is again not a participant. It flips the received bit, measures it, and records the result as \gemo{3} before sending \gemb{3} to \node{4}. 
The node \node{4} now acts as the first participant \node{2}: apply $C^{4}$, compute \gemb{4}, draw \gemo{4} and send \gemb{4} to \node{5}. 
This last participant \node{5} is now in a special position at the end of the network, which allows them to skip the last step and just to apply $C^{5}$, to compute \gemb{5} and to draw \gemo{5}.

\subsection{Measurements and post-processing}
\label{subsec:measure_post} The participants now use a fraction $p$ of the $L$ \GHZ{3} states to check for eavesdropping and the rest to generate conference keys. Using $L\cdot h_{2}(p)$ bits of the pre-shared key -- where \begin{equation*}
h_{2}(p) \defeq -p \log_{2}(p) - (1-p) \log_{2}(1-p)
\end{equation*}
is the binary entropy of $p$ -- Alice, Bob, and Charlie coordinate their measurements to be in either the $\sigma_{z}$-basis (for \texttt{KeyGen} rounds) or $\sigma_{x}$-basis (for \texttt{Verification} rounds). During the latter, $\langle m\rangle\defeq L\cdot p$ states are measured to estimate the $\sigma_{x}$-basis error rate
\begin{equation*}
Q_{X}^{m} \defeq \frac{1}{2}\left(1-\langle \sigma_{x}^{\qub{a}} \sigma_{x}^{\qub{b}}\sigma_{x}^{\qub{c}}\rangle\right),
\end{equation*}
that is the relative number of erroneous (\ie odd-parity) measurements.
Every party announces a uniformly random bit after every \texttt{KeyGen} and every \texttt{Verification} round -- with the exception of Bob and Charlie who announce their measurement outcome after each \texttt{Verification} round; this allows Alice to calculate $Q^{m}_{X}$. When Alice determines that $Q^{m}_{X}$ is above a predetermined tolerance threshold $Q_{\mathrm{tol}}$, she sets her \textit{abort bit} to $1$ to abort the protocol.

The other $\langle k \rangle \defeq L- \langle m \rangle=L\cdot\left(1-p\right)$ states are used to generate conference keys by Alice, Bob, and Charlie measuring in the $\sigma_{z}$-basis. This results in $k$ bits of raw key for each participant which is then post-processed with \textit{error correction} and \textit{privacy amplification}.

To perform error correction, Alice applies a publicly known error-correcting code to her raw key and encrypts the resulting error syndrome with a one-time pad using a portion of the pre-shared key. Alice now announces the encrypted syndrome, while everyone else announces a string of random bits of the same length. Bob and Charlie decrypt Alice's error syndrome using their pre-shared key and then perform error correction on their respective keys using the publicly known error-correcting code.

To verify the error correction, all participants apply a hash function $h_{\mathrm{EC}}$ to their corrected key; Alice announces her outcome after encrypting it with a one-time pad using part of the pre-shared key, while everyone else announces a uniformly random bit string instead. Bob and Charlie now both decrypt Alice's announced output and verify the error correction by comparing their output of $h_{\mathrm{EC}}$ with Alice's. If either Bob or Charlie find a discrepancy, they abort by setting their \textit{abort bit} to $1$; this ensures that the key they share is correct.

To potentially abort the protocol, the participants each announce their \textit{abort bit}, which is $1$ if and only if they want to abort, while the non-participants all announce random bits. The participants encrypt again by one-time padding their bit each with a separate single bit of the pre-shared key, using another $3$ bits in total.

Finally, the participants perform \textit{privacy amplification} to remove any correlation between the key and a potential eavesdropper, thereby reducing the length of the key. To this end, they apply another hash function $h_{\mathrm{PA}}$; after subtracting the necessary key to replenish the pre-shared key for communication in subsequent rounds, this results results in a secure and correct key $s$ of length $\ell$ (see App.~\ref{append:security}) shared by Alice, Bob, and Charlie if the protocol was not aborted.

\section{Analysis of the protocol}
\label{sec:analysis}
Our nearest neighbour state preparation ensures that no central server is needed to provision resources. First, the parties create the linear cluster state by exchanging Bell pairs with their nearest neighbours, performing an entangling operation between their two qubits and subsequently measuring one. The non-participants then measure their remaining qubit in order to establish a \GHZ{3} state between the participants.\footnote{Note that the $CZ$ operation and the two subsequent measurements of each non-participant can be viewed as a measurement in the Bell-state basis, \ie as a \textit{Bell-state projection}, and therefore can be performed in one step.} The resulting corrections to obtain the \GHZ{3} are non-trivial only for Alice and Charlie: The correction for Alice depends only on the number of non-participants between Alice and Bob and the measurement outcomes $\{\gemo{i}\}_{i=a+1}^{b-1}$ of these nodes, while the correction for Charlie can be constructed by analogy.
Importantly, the part of the correction that depends on the measurement results generates only Pauli corrections, so all these corrections can be considered as post-processing of \texttt{KeyGen} and \texttt{Verification} rounds of Alice and Charlie; this means that the participants do not have to wait for announcements from non-participants before performing their measurements.

\subsection{Anonymity of the protocol}
\label{subsec:anonymityanalysis}
The definition of anonymity is taken from \cite{grasselliSecureAnonymousConferencing2022}. The protocol is defined to be \textit{$\varepsilon_{\mathrm{an}}$-anonymous} if, for any choice of participants, it is at most $\varepsilon_{\mathrm{an}}$-close (in trace distance) to any state with a desired property. This property, adapted to our attack model, loosely states that the reduced state for \textit{any} subset $G \subset \net$ that does not contain the participants, is independent of the choice of participants \partic{}.
For the rigorous definition and more details, see App.~\ref{append:anonymity}.

The alternating $\sigma_{x}$-$\sigma_{y}$ measurement pattern of Step 2a in Protocol~\ref{prot:ghzextraction} only works when Alice and Charlie are at the respective ends of the linear cluster state. It is for this reason that we designed our extraction of the \GHZ{} state to effectively only use the middle state \Lin{\mathrm{m}}. 
Since all non-participants perform steps that are independent of the position of Alice, Bob, and Charlie, the nodes $\{\node{1},\dots, \node{a-1}\}$ and $\{\node{c+1},\dots, \node{n}\}$ generate the linear cluster states \Lin{\mathrm{l}} and \Lin{\mathrm{r}} as a byproduct. 
Alice and Charlie not performing the Bell projection creates the tri-separable network state
$
 \ket{\net} \defeq \Lin{\mathrm{l}} \otimes \Lin{\mathrm{m}} \otimes \Lin{\mathrm{r}}.
$
While this could make their actions distinguishable from those of all other nodes, it is easy to see that the reduced quantum state of each node \node{i} is maximally mixed -- even given all announced measurement results. 
In particular, this means that the reduced state of each node does not contain information about the identity of any other node in the network.

For noiseless scenarios, these announced measurement results $\{\spmo{i}~|~\node{i} \in \net \setminus \{\node{a}, \node{c},\node{1}\}$ are all uniformly random and uncorrelated -- see~App.~\ref{append:anonymity} for a proof. To mask their identity, the other three nodes therefore announce uniformly random bits \spmo{a}, \spmo{c} and \spmo{1}. For noisy scenarios, the announced measurement bits remain uniformly random, if there is no noise bias in the $\sigma_{x}$ or $\sigma_{y}$ basis, as for \eg \textit{depolarizing noise}.
If there were such a bias, it could however reveal some information about the participants' identities, as the announced measurement results would be distinguishable from the uniformly random announcements of the participants. This can be avoided if the latter introduce some bias to their source of randomness, and hence mimic the non-uniformity of the announced measurement results of the non-participants; see Sec.~\ref{sec:discussion} for more discussion.
Furthermore, the neighbours to-the-right of the participants are not aware that the latter announce randomly chosen bits and will therefore perform $Z$ corrections conditioned on them (\ie Step $1$ in Protocol~\ref{prot:statepreparation}); hence Step $2b$ of the protocol works as \say{anti-correction} performed by \Alice{}, \Charlie{} and \node{1} on their part of the Bell pair.

Similar to Protocol~\ref{prot:statepreparation}, during Protocol~\ref{prot:ghzextraction} all non-participants \nonpart{} (except $\node{1}$ and $\node{n}$) measure and announce their outcomes $\{\gemo{i} ~|~\node{i} \in \nonpart{}, i \notin \{1, n\}\}$, which are uniformly random and uncorrelated -- see~App.~\ref{append:anonymity} for a proof. To hide their identity, the participants \partic{} as well as $\node{1}$ and $\node{n}$ announce a uniformly randomly drawn bit (in case of biased noise, the same considerations and solution as in Protocol~\ref{prot:statepreparation} apply).

In approaches where anonymity is of no concern, an error syndrome can be announced publicly.
However, this syndrome might not be uniformly random and could potentially disclose Alice's identity. Hence Alice one-time-pad encrypts the error syndrome so that her communication is indistinguishable from all other parties; the same reasoning applies to all other communication of the post-processing (\ie the verification and abort), too.

\subsection{Security and performance of the protocol}
Following \cite{murta2020quantum}, we define the protocol to be $\varepsilon_{\mathrm{c}}$-correct if the probability that the generated keys are the same for all participants is greater than $1-\varepsilon_{\mathrm{c}}$. Similarly, we define the protocol to be $\varepsilon_{\mathrm{s}}$-secret if the output state is $\varepsilon_{\mathrm{s}}$-close (in trace distance) to an ideal state where the key is uniformly random and uncorrelated. The protocol is then called $(\varepsilon_{\mathrm{c}} + \varepsilon_{\mathrm{s}})$-secure. For more details and the rigorous definition, see App.~\ref{append:security}.

In principle, all resulting \GHZ{3} states could be used for key generation by each participant measuring in the $\sigma_{z}$ basis. However, to ensure secrecy of the key, it is of utmost importance that the states are verified, which is achieved by having all participants measure their qubit in the $\sigma_{x}$ basis instead. Crucially, the \texttt{Verification} rounds are selected such that potentially malicious non-participants do not learn of their selection. This is achieved by coordinating the \texttt{Verification} rounds in advance between Alice, Bob, and Charlie using secret communication. There are $m = L \cdot p$ \texttt{Verification} rounds and we therefore need $L \cdot h_{2}(p)$ bits to coordinate their choices.
As the $k$ bits of the raw conference key resulting from the $L \cdot (1-p)$ \texttt{KeyGen} rounds are neither perfectly correlated nor secret, \textit{error correction} and \textit{privacy amplification} are required. 

\paragraph{Error correction.}
To make the key $\varepsilon_{\mathrm{c}}$-correct they use a publicly known error-correcting code (e.g.,~a \emph{low-density parity-check} code as in Ref.~\cite{proietti_experimental_2020}) with an error syndrome of length $l_{\mathrm{EC}} \defeq k \cdot h_{2} (Q_{z})$. Here, $Q_{z} = \max_{i=B,C} \frac{1}{2}(1-\langle \sigma_{z}^{a}\sigma_{z}^{i}\rangle)$ is the maximum pairwise $\sigma_{z}$-basis error rate between Alice and Bob or Alice and Charlie; it is estimated in advance and thus regarded as a given parameter. Since the error syndrome is one-time padded, the participants use up $l_{\mathrm{EC}}$ of the pre-shared key. To verify the error correction, the participants apply a hash function $h_{\mathrm{EC}}$, which is drawn from a family of two universal hash functions using a seed $s_{\mathrm{hEC}}$. This seed is sourced from the pre-shared key\footnote{It should be noted that unlike the other uses of the pre-shared key, the same seed can be used in subsequent runs of the protocol and therefore it only needs to be determined once.\label{doublefootnote}} and has length $k - \log_{2}(\varepsilon_{\mathrm{c}}) - 1$, which ensures an $\varepsilon_{\mathrm{c}}$-correct key. 
The output of the hash function $h_{\mathrm{EC}}$ is a bitstring of length $\log_{2}(\frac{2}{\varepsilon_{\mathrm{c}}})$; the same amount is used from the pre-shared key to encrypt the outcome using one-time pad before Alice announces it.

\paragraph{Privacy amplification.}
Privacy amplification works in a similar fashion: Alice, Bob and Charlie apply a hash function $h_{\mathrm{PA}}$ drawn from a family of two-universal hash functions using a seed $s_{\mathrm{hPA}}$ of length $k + l_{\mathrm{PA}} - 1$ sourced from the pre-shared key$^\text{\ref{doublefootnote}}$,
where $l_{\mathrm{PA}} \defeq k \cdot \left(1 - h_{2}(Q_{\mathrm{tol}} + \mu\left(\frac{\varepsilon_{\mathrm{s}} - \varepsilon}{2}\right))\right) + 2 - 2\log_{2}(\varepsilon)$ is the length of the output of the hash function. Here, $\mu$ is a statistical correction and $\varepsilon_{\mathrm{s}}$ is the \textit{security} parameter, while $\varepsilon>0$ is a free parameter (see~App.~\ref{append:security} for details). Note that in other approaches \cite{grasselli_finite-key_2018}, privacy amplification also affects the $l_{\mathrm{EC}}$-parity bits that are transmitted during error correction. Since these have been encrypted with the pre-shared key in our approach, no leakage is possible and thus no privacy amplification is needed for these bits.

\begin{figure}[h]
 \centering
 \includegraphics[width=0.9\textwidth]{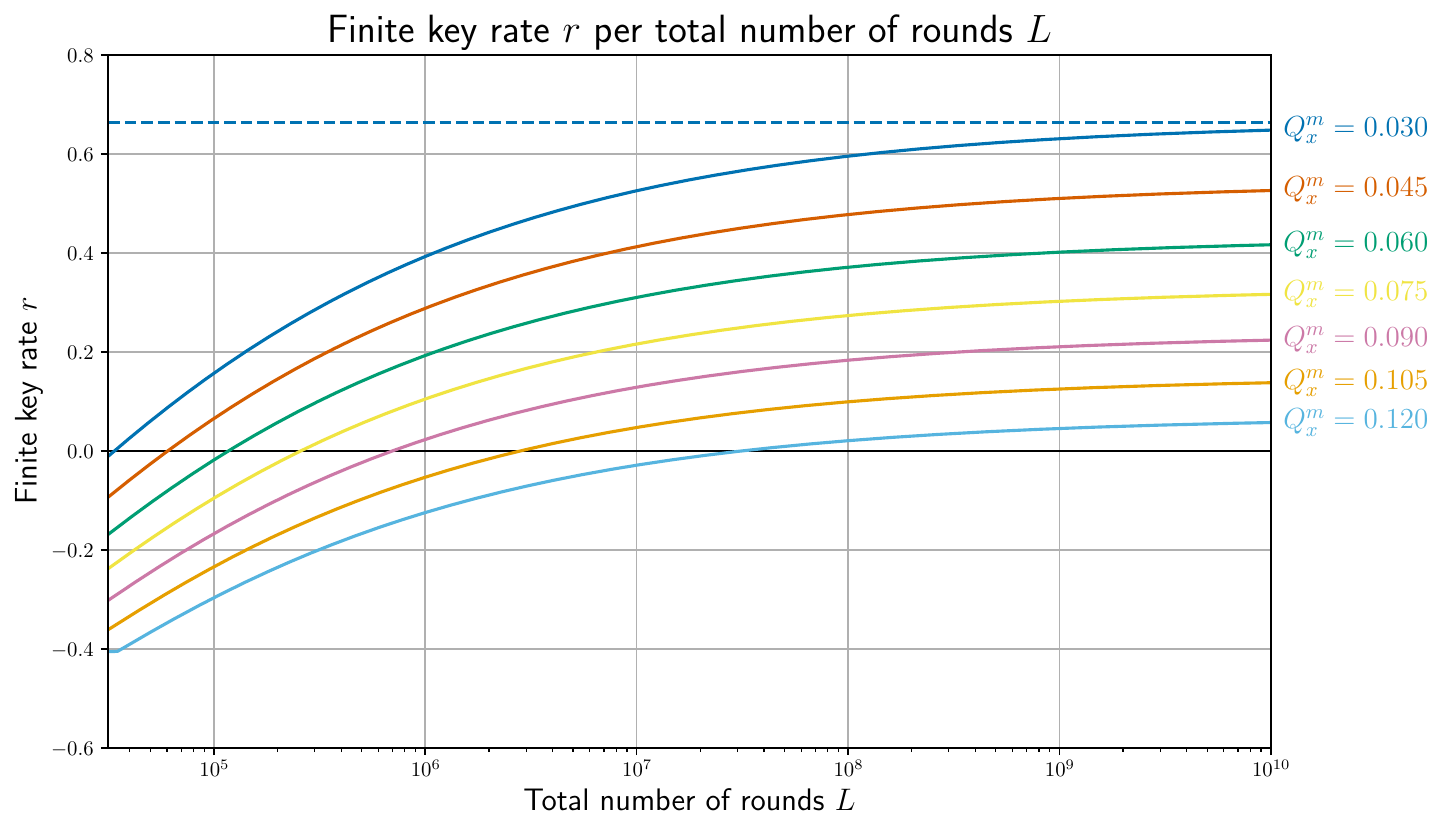}
 \caption{Finite key rate $r$ as a function of the total number of network states $L$. Here, $Q_\mathrm{tol}$ is fixed at various values of $Q_{x}^{m}$ and $Q_{z}$ is fixed at two thirds of this value to simulate white noise. The blue dotted line is the asymptotic key rate for the minimum $Q_{x}^{m}$ shown.
 Security parameters $\varepsilon_{\mathrm{c}}$ and $\varepsilon_{\mathrm{s}}$ have both been fixed at $10^{-8}$; the rates are optimised over $p$ and $\varepsilon$.}
 \label{fig:fkrperL}
\end{figure}

\paragraph{}
Finally, the unencrypted announcement of the -- not uniformly random -- \textit{abort} bits could reveal the identities of the participants. This is solved by all participants encrypting their abort bits with one-time pads; each participant uses a single bit of their pre-shared key to hide the correlation of the abort bits. In order to obtain an accurate key rate, we need to replenish the pre-shared key used as a one-time pad in the various communications: coordinating the verification rounds, the error correction's syndrome and its verification, and the three bits for the abort communication, obtaining the secret key length $\ell$ (see App.~\ref{append:security}). Dividing by the number of rounds $L$, we obtain the \textit{key rate} $r \defeq l/L$ 
with
\begin{equation*}\label{equation:fkr}
 r = (1-p)\left[1 - h_{2}\left(Q_{\mathrm{tol}} + \mu \left(\frac{\varepsilon_{\mathrm{s}} - \varepsilon}{2}\right)\right) - h_{2}(Q_{z})\right] - h_{2}(p) +\frac{1}{L}
 \left(
 \log_{2}\left(\varepsilon^2 \varepsilon_{c}\right)
 -2
 \right),
\end{equation*}
where $\mu\left(\frac{\varepsilon_{\mathrm{s}}-\varepsilon}{2}\right)$ is a statistical correction (see App.~\ref{append:security}) and $\varepsilon>0$ is a free parameter. Note that for fixed $ \varepsilon_\mathrm{s}>0$ and $\varepsilon_\mathrm{c}>0$ and given $Q_\mathrm{tol}$, $Q_{z}$ and $L$, one can optimise over $p$ and $\varepsilon$. In the asymptotic limit (\ie $L \rightarrow \infty$), not only the $L$-dependent terms of Eq.~\eqref{equation:fkr} vanish, but so do $p$ and $\mu$. 
We refer to Tab.~\ref{tab:variousLforell} for a representation of the performance of the protocol in terms of the required number $L$ of network states for a fixed key length $\ell$, while Fig.~\ref{fig:fkrperL} contains the achievable finite key rate $r$ as a function of $L$.

% \vspace{-0.2cm}
\begin{framed}
\noindent\begin{minipage}[htpb]{0.62\textwidth}
\includegraphics[width=\textwidth]{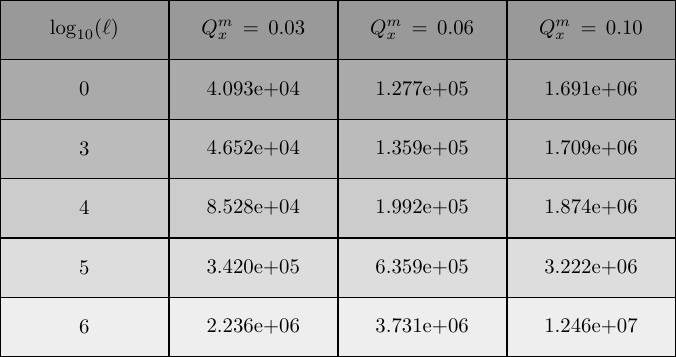}
\end{minipage}
\hspace{0.01\textwidth}
\begin{minipage}[t!]{0.35\textwidth}
 \vspace{0.35cm}
 \captionof{table}{
 The number of network states $L$ necessary to obtain various secret key lengths $\ell$, for different $\sigma_{x}$-basis error rates $Q_{x}^{m}$ ($Q_\mathrm{tol}$ has been fixed at this value). Here, $Q_{z}$ is fixed at two thirds of $Q^{m}_{x}$ to simulate white noise. The security parameters $\varepsilon_{\mathrm{c}}$ and $\varepsilon_{\mathrm{s}}$ have both been fixed at $10^{-8}$ and the rates are optimised over $p$ and $\varepsilon$.}
 \label{tab:variousLforell}
\end{minipage}
\hspace{0.01\textwidth}
\end{framed}

\section{Discussion}\label{sec:discussion}
In this work, we have investigated how to anonymously establish a secret key between three participants in a line of nearest-neighbour quantum nodes. We find that by sharing maximally entangled pairs that are then projected into linear cluster states, a secret key can be obtained without revealing the identity of the participants. This contrasts with previous approaches to anonymous conference key agreement \cite{hahn2020anonymous, grasselliSecureAnonymousConferencing2022}, which directly distribute large \GHZ{} states that are not only error-prone but also rely on a central server. 

Although in this work we explicitly show how the three-partite GHZ state is generated using linear cluster states, this is in fact equivalent to performing Bell-state projections, as is for example common in quantum repeater schemes. There are other ways to extract smaller GHZ states from Bell pairs \cite{meignant2019distributing}, however, these might not be as straightforward to perform anonymously. We also note that our method of generating the linear cluster state connecting the three participants, is not unique: in fact, any scheme that anonymously generates a linear cluster state from Alice to Charlie can be used, and from there, our proposed method for generating the conference key can be further applied. 

With respect to non-trusted participants, if they were allowed to collaborate with an adversary who controls the entanglement sources and does not care about protocol termination, a trivial attack would be to distribute an eigenstate of a to-be-measured observable to one or more of the parties. Since participants and non-participants behave differently and the protocol contains announcements, albeit encrypted ones, the attacker would, at least probabilistically, learn the roles of the parties in the protocol. This attack can in principle be circumvented by first performing randomized verification of the shared entanglement as in Ref.~\cite{unnikrishnan_anonymity_2019, hahn2020anonymous}; however, this would require access to sufficiently good quantum memories and/or would drastically decrease the keyrate. In the absence of such countermeasures, anonymity can only be guaranteed against the non-trusted participants assuming they cannot collaborate with an all powerful eavesdropper. However, we emphasise that the security, as opposed to the anonymity, does hold in a fully adversarial model where all non-trusted participants and the eavesdropper collaborate (see Appendix~\ref{append:security}).

As mentioned in Section~\ref{subsec:anonymityanalysis}, the measurement outcomes that are announced throughout the protocol could, in the presence of noise, be distinguishable from the chosen (uniformly random) announced bits of the nodes that do not perform a measurement; this would be detrimental to their anonymity, even if the non-trusted participants are not collaborating with an eavesdropper, since they could potentially learn about noise characteristics in the network via measurements announced in the current (or previous) runs of the protocol.
To mitigate this risk, these nodes can add a bias to their random announcements that is consistent with the noise of their detectors. If they want to mimic the \textit{exact} bias that their detectors have, they could -- in an adjusted protocol -- postpone all announcements until all $L$ rounds of measurements have taken place.
Such a protocol would be modified such that \node{1} and the participants can secretly estimate their bias, by performing additional measurements in the bases that they would have had to perform if it wasn't for their special role. After all measurements have taken place everyone can then announce their (potentially fabricated) bits. Since these secret measurements are in place of the actual measurements that need to be performed, the corresponding round cannot be used for either verification or key generation. As such, this approach would slightly diminish the keyrate.

We now briefly discuss the prospects for implementation of these protocols via multi-partite photonic entanglement. The dominant cause of noise is typically presence of higher-order photon events for parametric down conversion (PDC) photon sources \cite{thalacker_anonymous_2021,Wang:2016dk,proietti_experimental_2020,Ruckle:2022jg,Pickston:2022dw} and photon distinguishability for solid state sources \cite{Xiang:2020ks,Yang:2022cl,Thomas:2022js}, although other effects such as detector dark-counts and misalignment also contribute. Furthermore, for practical implementations it is important that the generated photons are at telecom wavelengths compatible with low loss transmission over optical fibre, which renders some multi-partite entanglement sources unsuitable. 

Regardless of the source of the noise, for the ACKA protocol the only quantity that ultimately matters are the measured QBER’s in the X and Z bases. In that sense our results in Fig.~\ref{fig:fkrperL} already provide a detailed description of the tradeoff between the total tolerable noise and block size. An implementation of our ACKA protocol has already been carried out demonstrating positive asymptotic keyrates, albeit at non-telecom wavelengths \cite{Ruckle:2022jg}. Comparison with other experiment results shows that for moderate block-sizes the required noise thresholds are well within those that have already been demonstrated in state-of-the-art PDC sources deployed through optical fibre at telecom wavelengths for 4 and 6 photon entangled states \cite{proietti_experimental_2020,Pickston:2022dw} showing that proof-of-principle demonstrations of this protocol are within the reach of present-day technology. Nevertheless, the observed rates are quite low due to losses, and these would only become more severe with increasing transmission distance and number of parties. This observation, coupled with previous work on the increasingly demanding noise thresholds for large scale multi-partite CKA \cite{murta2020quantum,epping_multi-partite_2017, grasselli_finite-key_2018, grasselli2019conference}, suggest that a robust, large-scale implementation would require more sophisticated networks incorporating quantum repeaters or error-correction protocols \cite{Pirandola:20,Roadmap, khatri2020principles}.

We leave the generalization of our protocol to more than three participants as an open question: the extraction of larger \GHZ{} states (\ie more than three qubits) from linear cluster states is possible (see Ref.~\cite{dejongExtractingMaximalEntanglement2022, hahn_quantum_2019}). However, the size of the larger \GHZ{} state is bounded from above by roughly half of the number of nodes between Alice and Charlie \cite{dejongExtractingMaximalEntanglement2022}. 
In addition, the specific measurement bases used to extract the state may depend on the position of the participants, so particular care must be taken to prevent identity leakage in this way. 
Finally, closed-form expressions for post-processing corrections in this generalized form are not straightforward either -- and obtaining them remains an open question.
Our work contributes to the growing body of multi-partite quantum cryptographic schemes that live up to stringent security requirements in protocols that go beyond point-to-point protocols. It is the hope that this work further stimulates 
theoretical and experimental research towards understanding notions of secure quantum communication in multi-partite settings.

\section{Acknowledgements}
We thank Stefanie Barz, Jakob Budde, Lukas Rückle and Christopher Thalacker for fruitful discussions during our ongoing collaboration on experimentally implementing our protocol. We thank Federico Grasselli for carefully reading an early version of the manuscript and providing valuable feedback. J.~de Jong and A.~Pappa acknowledge support from the Emmy Noether DFG grant No.~418294583. F.~H.~acknowledges support from the German Academic Scholarship Foundation and J.~E.~from the BMBF (Q.Link.X and QR.X), J.~E.~and A.~P.~ also acknowledge support from the Einstein Research Unit on Quantum Devices, for which this is an inter-node project.

\newpage
%%%%%%%%%%%%%%%%%%%%%%%%%%%%%%%%%%%%%%%%%%%%%%%%%%%%%%%%%%%%%%%%%%%%%%%%%%%%%%%%%%%%%%%%%%% APPENDICES %%%%%%%%%%%%%%%%%%%%%%%%%%%%%%%%%%%%%%%%%%%%%%%%%%%%%%%%%%%%%%%%%%%%%%%%%%%%%%%%%%%%%%%%%%%%%%
\appendix
\section*{Appendices}
The appendix consists of three parts. Part~\ref{append:corrections} contains an explanation of the corrections that the participants have to perform on their qubits due to the network layout and the measurement results of the non-participants. Part~\ref{append:security} contains a restatement of the protocol, the definition of security, and the proof of security for the protocol. Finally, part~\ref{append:anonymity} deals with the anonymity of the protocol, showing in particular that the announcements of the measurement do not reveal any information about the identity of the participants or non-participants.

\section{Corrections for Alice, Bob and Charlie during Protocol~\ref{prot:ghzextraction}}\label{append:corrections}
\begin{figure}[ht!]
\begin{framed}
 \centering
 \includegraphics[width=\textwidth]{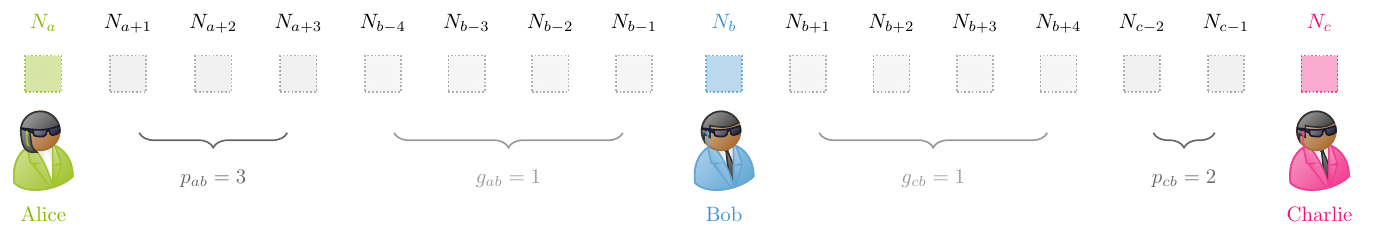}
 \includegraphics[width=\textwidth]{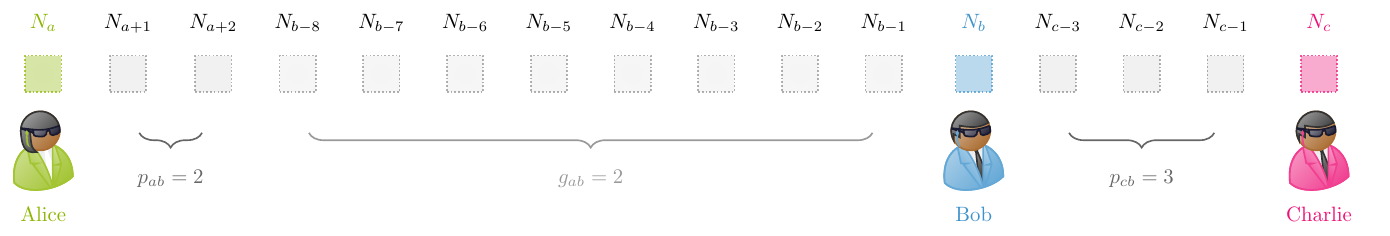}
 \caption{Two exemplary configurations. \textbf{Top:} $\delta_{ab}=7$ (with $\modfourval{} = 3$ and $\fourgroup{} = 1$) and $\delta_{cb} = 6$ (with $p_{cb} = 2$ and $g_{cb} = 1$). \textbf{Bottom:} $\delta_{ab}=10$ (with $\modfourval{} = 2$ and $\fourgroup{} = 2$) and $\delta_{cb} = 3$ (with $p_{cb} = 3$ and $g_{cb} = 0$).}
 \label{fig:correctionsexample}
\end{framed}
\end{figure}

Alice and Charlie need to perform a correction to obtain the \GHZ{3} state with Bob, whereas Bob never has to perform a non-trivial rotation. The corrections for Alice and Charlie are structurally similar; we first introduce those for Alice. In order to achieve this, we define the following quantities.
\begin{itemize}
 \item $\delta_{ab} \defeq b - a - 1$, the number of non-participants between Alice and Bob.
 \item $\modfourval \defeq \delta_{ab} \mod 4$, the mod-four value of $\delta_{ab}$
 \item $\fourgroup \defeq \frac{\delta_{ab} - \modfourval}{4}$, the integer number of groups of four that fit between Alice and Bob.
\end{itemize}
For Charlie, $\delta_{cb}, g_{cb}$ and $p_{cb}$ are defined in a similar fashion.
We refer to Fig.~\ref{fig:correctionsexample} for two potential configurations of the network that exemplifies these definitions.

\noindent Alice now performs the following correction steps:
\begin{enumerate}
 \item Apply a \textit{configuration} correction $C_{ab}$ depending on \modfourval{} and \fourgroup, as shown in Tab.~\ref{tab:corrections}, picking the left (\textcolor{brown}{brown}, $\beta_{a}=1$) or right (\textcolor{green}{green}, $\beta_{a}=0$) table.
 \item Divide all the measurement outcomes $\{m_{i}\}_{a+1}^{b-1}$ into a set $\{m_{i}\}_{a+1}^{a+1+\modfourval}$ and a set $\{m_{i}\}_{a+2+\modfourval}^{b-1}$ -- where it is to be understood that if $\modfourval = 0$, the first set is empty.
 \item From the outcomes in the first set, they calculate the bits $k_{x}$ and $k_{z}$ using Tab.~\ref{tab:corrections}.
 \item From the outcomes in the second set, out of every pair of four they select the measurement outcomes as described in Tab.~\ref{tab:Corrections_lx_lz} and add them all together to calculate $l_{x}$ and $l_{z}$, respectively (\eg if $\beta_{a}=1$, Alice selects every odd element of the second set to calculate $l_{x}$, and every second, third and fourth out of four to calculate $l_{z}$).
 \item They apply an $X$ operation on their qubit if and only if $k_{x} \oplus l_{x} = 1$.
 \item They apply a $Z$ operation on their qubit if and only if $k_{z} \oplus l_{z} = 1$.
\end{enumerate}

\begin{table}[ht!]
\centering
 \includegraphics[width=0.49\textwidth]{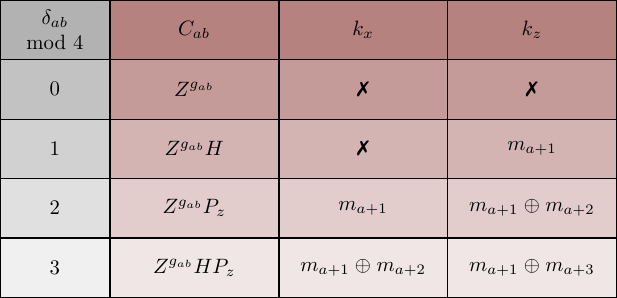}
 \includegraphics[width=0.49\textwidth]{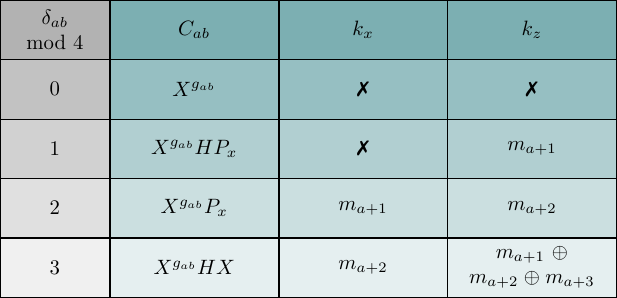}
 \caption{Local corrections that Alice needs to perform to obtain the \GHZ{} state with Bob and Charlie after the non-participants measured their qubits. The \textcolor{brown}{left table} shows the corrections if the non-participant $a+1$ after Alice measured \textcolor{brown}{in the $\sigma_{x}$-basis ($\beta_{a}=1$)}, the \textcolor{green}{right table} the corrections if it was \textcolor{green}{in the $\sigma_{y}$-basis ($\beta_{a}=0$)}. \textbf{The $\mathbf{C_{ab}}$ column} contains the \textit{configuration} correction which only depends on the number of non-participants $\delta_{ab}$ between Alice and Bob -- note that $\fourgroup \defeq \lfloor {\delta_{ab}}/{4} \rfloor$. \textbf{The $\mathbf{k_x}$ column} contains the measurement outcomes that add to $k_{x}$, which induce together with $l_{x}$ a correction $X^{k_{x} \oplus l_{x}}$; similarly \textbf{the $\mathbf{k_z}$ column} contains the measurement outcomes that create $k_{z}$.}
 \label{tab:corrections}
\end{table}

\begin{wraptable}{r}{0.33\textwidth}
 \begin{center}
 \vspace{-.8cm}
 \includegraphics[width=0.33\textwidth]{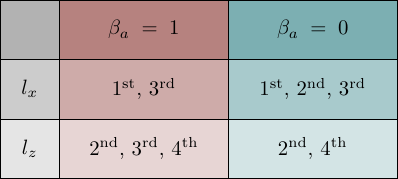}
 \end{center}
 \vspace{-0.cm}
 \caption{Selection of measurement outcomes out of every pair of four from the second set to calculate $l_{x}$ and $l_{z}$, respectively. For example, when $\delta_{ab} = 7$ and $\gemb{a}=1$, $l_{x} = m_{a+ 4} \oplus m_{a + 6}$ and $l_{z} = m_{a + 5} \oplus m_{a+6} \oplus m_{a+7}$.}
 \label{tab:Corrections_lx_lz}
\end{wraptable} 
Note that all three corrections (\ie the configuration correction, the $X$ correction and the $Z$ correction) can be contracted into a single Clifford operation. However, since the measurement-outcome dependent corrections are only Pauli operators, they will at most flip the measurement outcomes for Alice in the subsequent steps of the protocol -- and need not be physically implemented. 
This also means that the participants can perform their \texttt{KeyGen} or \texttt{Verification} measurements \textit{before} the measurement outcomes of the non-participants are announced. By having all nodes $\{\node{a+1},\dots, \node{b-1}\}$ perform their measurements and Alice subsequently perform the aforementioned corrections, the linear cluster state is contracted towards a $\Lin{a,b,b+1 ,\dots, c-1,c}$ linear cluster state. Hence, Charlie can perform the same steps (while using the measurement outcomes $\{m_{c-1}, m_{c-2} ,\dots, m_{b+1}\}$, $\delta_{bc} \defeq c - b - 1$ and its redefined derivatives) to contract the state towards a three-partite linear cluster state \Lin{a, b, c}. Two final $H$ gates for Alice and Charlie result in the desired \GHZ{3} state between Alice, Bob and Charlie.

\subsection{Calculating the corrections}
Using the stabiliser formalism, it is straightforward to show that, starting from a linear cluster state \Lin{a,a+1,\dots, c-1,c}, a measurement on node \node{a+1} in the $\sigma_{x}$- or $\sigma_{z}$-basis results in \Lin{a,a+2,\dots, c-1,c} up to a local correction $C_{ab}^{a+1}$ for Alice, where this correction depends on both the measurement basis $\beta_{a+1}$ and outcome $m_{a+1}$ as
\begin{align}
 C_{ab}^{a+1}(m_{a+1},\beta_{a+1}) = P_{z}^{(2m_{a+1}+\beta_{a+1})}H = H P_{x}^{(2m_{a+1}+\beta_{a+1})},
\end{align}
where $P_{z} \defeq R_{z}\left(\frac{\pi}{2}\right)$ is a half-rotation around the $Z$-axis and $P_{x}$ is defined similarly. Note that either identity (\ie the $Z$- or $X$-based rotation) can be used.

A series of multiple measurements then introduces a concatenation of these corrections, where the corrections are performed in order from \node{a+1} to \node{b-1}. They do not necessarily commute, but by using the $X$- and $Z$-based correction interchangeably (and thus cancelling out the $H$ operations), and using the identity $Z^{m_{i}}P_{x} = P_{x}X^{m_{i}}Z^{m_{i}}$ (and likewise for $P_{z}$) one can group all the corrections that are not measurement outcome based together as the first corrections; this allows to partition the complete correction into a `configuration' correction and an outcome-based correction.

Specifically, for the alternating pattern of $\sigma_{x}$ and $\sigma_{y}$ measurements, each group of four consecutive measurements together introduces only Pauli corrections. For example, for any group of four consecutive nodes $\{\node{1},\node{2},\node{3},\node{4}\}$ (note that these labels resemble \textit{any} set of four consecutive nodes) these corrections are
\begin{align*}
& X^{(m_{1}+m_{3})}Z^{(m_{2}+m_{3}+m_{4})}X, \tag{$\beta_{a}=1$}\\
& X^{(m_{1}+m_{2}+m_{3})}Z^{(m_{2}+m_{4})}Z. \tag{$\beta_{a}=0$}
\end{align*}
Up to an irrelevant global phase, all these operators commute with each other. Therefore, starting from the last measured node (\ie \node{b-1}) an integer multiple of four can be `stitched together'. Since there are $\fourgroup{} \defeq \lfloor {\delta_{ab}}/{4} \rfloor$ of such groups, the correction becomes
\begin{align*}
\tag{$\beta_{a}=1$} & \prod_{i=0}^{g_{ab}-1}X^{\left(m_{b-4i-4} \oplus m_{b-4i-2}\right)}Z^{\left(m_{b-4i-3} \oplus m_{b-4i-2} \oplus m_{b-4i-1}\right)}X = X^{l_{x}} Z^{l_{z}} X^{\fourgroup}, \\
\tag{$\beta_{a}=0$} & \prod_{i=0}^{g_{ab}-1}X^{\left(m_{b-4i-4} \oplus m_{b-4i-3} \oplus m_{b-4i-2}\right)}Z^{\left(m_{b-4i-3} \oplus m_{b-4i-1}\right)}X = X^{l_{x}} Z^{l_{z}} Z^{\fourgroup},
\end{align*}
where $l_x$ is defined as
\begin{align*}
\tag{$\beta_{a}=1$} l_{x} \defeq & \bigoplus_{i=0}^{g_{ab}-1} m_{b-4i-4} \oplus m_{b-4i-2},\\
\tag{$\beta_{a}=0$} l_{x} \defeq & \bigoplus_{i=0}^{g_{ab}-1} m_{b-4i-4} \oplus m_{b-4i-3} \oplus m_{b-4i-2},
\end{align*}
and $l_z$ is defined as
\begin{align*}
\tag{$\beta_{a}=1$} l_{z} \defeq & \bigoplus_{i=0}^{g_{ab}-1} m_{b-4i-3} \oplus m_{b-4i-2} \oplus m_{b-4i-1},\\
\tag{$\beta_{a}=0$} l_{z} \defeq & \bigoplus_{i=0}^{g_{ab}-1} m_{b-4i-3} \oplus m_{b-4i-1}.
\end{align*}
The corrections for the measurements of the nodes $\node{a+1},\dots, \node{a+p_{ab}}$ (\ie the first \modfourval{} measurements) are then also grouped together; by splitting them into a measurement-outcome dependent and -independent part, they can be written as $X^{k_{x}}Z^{k_{z}}C_{ab}$, where $C_{ab}$, $k_{x}$ and $k_{z}$ can be read from Tab.~\ref{tab:corrections}. Note that the $X^{\fourgroup}$ or $Z^{\fourgroup}$ in Tab.~\ref{tab:corrections} is technically not part of the correction here, but that they will commute with $X^{k_{x}}Z^{k_{z}}$ and hence the total correction that Alice needs to perform becomes (where now $C_{ab}$ is as in Tab.~\ref{tab:corrections}):
\begin{equation}
 X^{l_{x}}Z^{l_{z}}X^{k_{x}}Z^{k_{z}}C_{ab} ~\hat{=}~X^{k_{x} \oplus l_{x}}Z^{k_{z} \oplus l_{z}}C_{ab},
\end{equation}
where $\hat{=}$ here indicates `up to an (irrelevant) global phase'.
Since these corrections only consider nodes between Alice and Bob, and since there are actions that Bob needs to perform, the corrections for Charlie work in a similar fashion and can be seen separately from these.

\section{Protocol statement and security proof}
\label{append:security}
We now state the protocol and proof the security of the generated key. Note that we have omitted the network state generation (\ie Protocol~\ref{prot:statepreparation}) as it does not affect security.

\paragraph{Input:}
\begin{itemize}
 \item $L$ network states $\ket{\net}$ connecting $\{\node{i}\}_{i=1}^n$, including \Alice, \Bob ~and \Charlie.
 \item Desired secrecy parameter $\varepsilon_\mathrm{s}>0$, which determines a correlation threshold $Q_{\mathrm{tol}}$, and correctness parameter $\varepsilon_\mathrm{c}>0$.
 \item A random string $s_\mathrm{b}$ of length $L\cdot h_2(p)$ secretly pre-shared between the participants to randomly choose $m$ out of the $L$ cluster states to be measured in the $\sigma_x$-basis for parameter estimation where $p=m/L$, leaving $k \defeq L-m$ measurements in the $\sigma_{z}$-basis for key generation.
 \item An estimate of the expected bit error rate $Q_{z}$ in the $\sigma_z$-basis between Alice and Bob and Alice and Charlie. The worst of these will be used to select an error-correcting code that requires an error syndrome of length $\ell_{\mathrm{EC}} \defeq k \cdot h_{2}(Q_{z})$ to be announced. 
 \item A pre-shared secret random string $s_{\mathrm{EC}}$ of length $\ell_{\mathrm{EC}}$ to be used to one-time pad the error reconciliation announcements, another pre-shared string $s_{\mathrm{hEC}}$ of length $\ell_{\mathrm{hEC}} \defeq \log_{2}(2/\varepsilon_\mathrm{c})$
 to one-time pad the error correction verification announcements, and three bits of pre-shared key to communicate aborting by the participants.
 \item Two pre-shared random strings, $s_{\mathrm{h}}$ and $s_{\mathrm{hEC}}$, of lengths $k + \ell_{\mathrm{PA}} - 1$ and $k + \ell_\mathrm{hEC} - 1$ respectively to be used as the seeds for hashing, where $\ell_\mathrm{PA}$ is the output of the privacy amplification hashing as defined below. The string $s_\mathrm{h}$ is used for privacy-amplification of the private key, while $s_{\mathrm{hEC}}$ is used to verify the error correction step has succeeded. Note that unlike the previous seeds, these can be used indefinitely and need not be replenished after each run of the protocol.
 \end{itemize}

\paragraph{Output:}
A key of length $\ell$ shared anonymously between Alice, Bob and Charlie that is $\varepsilon_\mathrm{s}$-secret and $\varepsilon_\mathrm{c}$-correct.
\begin{enumerate}
 \item For $i=1,\ldots,n$:
 \begin{enumerate}
 \item Node $\node{i}$ receives bit $\gemb{i-1}$ and computes $\gemb{i} = 1-\gemb{i-1}$, except for \node{1} who draws a random bit $\gemb{0}$ instead.
 \begin{enumerate}
    \item If $\node{i} \in \nonpart$, they measure the operator $\sigma_{x}^{i}$ or $\sigma_{y}^{i}$ if $\gemb{i} = 0$ or $\gemb{i}=1$, respectively. They broadcast the measurement outcome $\gemo{i}$.
    \item If $\node{i} \in \partic{}$, they announce a uniform randomly drawn bit $\gemo{i}$.
 \end{enumerate}
 \item Node $\node{i}$ sends bit $\gemb{i}$ to neighbour $\node{i+1}$, except for node \node{n}.
 \end{enumerate}
 \item The participants perform corrections on their qubits to rotate the post-measurement state to the desired \GHZ{} state.
 \begin{enumerate}
 \item Alice and Charlie apply their \textit{configuration corrections} $C_{a}$ and $C_{c}$, respectively (cf.~Tab.~\ref{tab:corrections}).
 \item Alice $(i=a)$ and Charlie $(i=c)$ both calculate their parameters $l^{i}_{x}, k^{i}_{x}$ and $l^{i}_{z}, k^{i}_{z}$ from the measurement outcomes of the non-participants (cf.~Tabs.~\ref{tab:corrections} and~\ref{tab:Corrections_lx_lz}) and each apply $X^{l_{x} \oplus k_{x}}$ and $Z^{l_{z} \oplus k_{z}}$ to their qubit.
 \item Alice and Charlie each apply a Hadamard operation $H$ to their qubit to obtain the final desired \GHZ{} state.
 \end{enumerate}
 
 \item Using the pre-shared string $s_{b}$, the participants coordinate their measurements of all $L$ \GHZ{} states into $m$ \texttt{Verification} rounds (\ie $\sigma_{x}$-basis) and $k$
 \texttt{KeyGen} rounds (\ie
 $\sigma_{z}$-basis). Everyone announces after each measurement a random bit $m_{i}$, except for Bob and Charlie, who announce their measurement outcomes for the \texttt{Verification} rounds.
 \item Alice, who can locate Bob's and Charlie's measurement outcomes from the \texttt{Verifi-} \texttt{cation} rounds, estimates the $\sigma_{x}$-basis error rate $Q_{X}^{m}=\frac{1}{2}(1-\langle \sigma_{x}^{\qub{a}} \sigma_{x}^{\qub{b}}\sigma_{x}^{\qub{c}}\rangle)$. If this is above the \textit{tolerance} $Q_{\mathrm{tol}}$, she aborts by setting her \textit{abort bit} to $1$.
 
 \item Alice computes the necessary information for error correction -- the \textit{error syndrome} of length $\ell_{\mathrm{EC}}$ -- and then one-time pad encrypts this information with the string $s_{\mathrm{EC}}$. All other players announce uniform random strings of length $\ell_{\mathrm{EC}}$.
 
 \item Bob and Charlie use their copies of $s_{\mathrm{EC}}$ to obtain $l_{\mathrm{EC}}$ and correct their $k$ $\sigma_z$ measurement strings, \ie their raw key. Alice, Bob and Charlie then hash their string using the seed $s_{\mathrm{hEC}}$. Alice encrypts her output using her copy of $s_{\mathrm{hEC}}$. Using their copy, Bob and Charlie each decrypt Alice's hash outcome and compare it to their own; if they do not align, they abort by setting their \textit{abort} bit to $1$.
 
 \item Alice, Bob and Charlie, using another three bits of the pre-shared key, encrypt their \textit{abort bit} -- which is equal to $1$ if and only if they want to abort -- and announce it, while all other parties announce uniformly random bits instead. If any participants announced a $1$, everyone aborts (meaning they will not use the generated key).
 
 \item Finally, the participants hash their measurement results with the seed $s_{\mathrm{hPA}}$ to produce the final key $s$ of length 
 \begin{equation*}
 l_{\mathrm{PA}} \defeq k \left[1 - h_{2}(Q_{\mathrm{tol}} + \mu\left(\frac{\varepsilon_{\mathrm{s}} - \varepsilon}{2}\right))\right] + 2 + 2\log_{2}(\varepsilon) = \ell + \ell_{\mathrm{EC}} + \ell_{\mathrm{hEC}} + L\cdot h_2(p) + 3.
 \end{equation*}
 However, to fairly evaluate performance the parties should replenish their stock of secret-shared key so as to be able to perform subsequent CKA protocols. Subtracting off the non-reusable shared randomness results in a string of length $\ell$ that is available for applications.
\end{enumerate}

We now prove the security of our protocol in the scope of an even more general adversary model than the one introduced in the main text, so that we can resort to a powerful machinery that has been developed in the literature \cite{Tomamichel:2012ffa,Portmann:2021} and we can build on the strategy of proof laid out in Ref.~\cite{grasselli_finite-key_2018}; the security of our protocol within our adversary model then follows readily.
However, there are some variations to the tools necessary to preserve the anonymity of the participants
which is key to the present work. We briefly explain some critical quantities and definitions. Let $\rho_{S_{A} S_{B} S_{C} E'}$ be the joint classical-quantum state between the final keys of the participants and an eavesdropper conditioned on passing all checks. Note that the eavesdroppers system, $E' = ER$, is made up of a quantum system, $E$, that completely purifies the pre-measurement state $\rho_{ABC}$ (and is, therefore, assumed to include system of the non-participating player) and a classical register $R$ that contains all of the information announced during the protocol. A protocol is called $\varepsilon_{\mathrm{rob}}$-robust if it passes the correlation and the error correction checks with probability $1-\varepsilon_{\mathrm{rob}}$. Defining a uniformly distributed state as
\begin{eqnarray}
\rho_{\mathbf{U}} \equiv \sum_{s \in \mathcal{S}} \frac{1}{|\mathcal{S}|} |s\rangle\langle s| 
\end{eqnarray}
with $\mathcal{S}$ the set of possible secret keys we have the following definition \cite{grasselli_finite-key_2018}.

\begin{definition}[Approximate robustness and secrecy]\label{QKDdef}
A CKA protocol that is $\varepsilon_{\mathrm{rob}}$-robust is $\varepsilon_c$-correct if
\begin{eqnarray}
\left(1-\varepsilon_{\mathrm{rob }}\right) \operatorname{Pr}\left[S_A\neq S_B \lor S_A\neq S_C\right] \leqslant \varepsilon_{c}
\end{eqnarray}
and $\varepsilon_s$-secret if
\begin{eqnarray}
\left(1-\varepsilon_{\mathrm{rob}}\right) \frac{1}{2}\left\|\rho_{S_{A}E'}-\rho_{\mathrm{U}} \otimes \rho_{E^{\prime}}\right\| \leqslant \varepsilon_{s} \label{es}
\end{eqnarray}
is called $(\varepsilon_s + \varepsilon_c)$-secure if it is $\varepsilon_c$-correct and $\varepsilon_c$ secret. \label{secdef}
\end{definition}
Turning first to multi-partite error correction we have the following statement.

\begin{theorem}[Theorem 2 in Ref.~\cite{grasselli_finite-key_2018}] \label{ecth}

Given a probability distribution $P_{X_A,B_1,B_2,\dots ,B_N}$, between Alice and $n$ other players there exists a one-way error-correction protocol for all $n$ players that is: $\varepsilon_{c}$-correct, and $2(n-1) \varepsilon^{\prime}$ -robust on $P_{X_A,B_1,B_2,\dots ,B_n},$ and has leakage
\begin{eqnarray}
\ell_{\mathrm{EC}} \leqslant \max_{i}
H_{0}^{\varepsilon^{\prime}}
\left(X_A|B_{i}\right)+\log_{2} \frac{2(n-1)}{\varepsilon_{c}}.
\end{eqnarray}
\end{theorem}

In terms of secrecy the critical results are leftover hashing against quantum side-information, an entropic uncertainty relation for smoothed min- and max-entropies, applied to our protocol, 
states the following. 

\begin{lemma}[Leftover hashing against quantum side information in Refs.~\cite{Tomamichel:2011ci,Tomamichel:2012ffa}] \label{lhl}
Let $\varepsilon'\geq 0$ and $\rho_{Z_AE}$ be a classical-quantum state where $Z_A$ is defined over a discrete-valued and finite alphabet, $E$ is a quantum system and $R$ is a register containing the classical information learnt by Eve during information reconciliation. If Alice applies a hash function, drawn at random from a family of two-universal hash functions that maps $\rho_{Z_AE}$ to $\rho_{S_AE}$ and generates a string of length $\ell$, then
\begin{eqnarray}
\frac{1}{2}\left\|\rho_{S_{A}ER}-\rho_{\mathrm{U}} \otimes \rho_{ER}\right\| \leqslant 2^{-\frac{1}{2} (\hmin^{\varepsilon'}(Z_A|ER) - \ell + 2)} + 2\varepsilon', \label{TD}
\end{eqnarray}
where $H_{\min }^{\varepsilon'}\left(Z_{A}|ER\right)$ is the conditional smooth min-entropy of the raw measurement data given Eve's quantum system and the leakage of the information reconciliation.
\end{lemma}
This leads to the following corollary.
\begin{corollary}[Secret string extraction] \label{corol1}
For an $\varepsilon_{\mathrm{rob}}$-robust protocol an $\varepsilon_s$-secret string of length
\begin{eqnarray}
\ell=H_{\min}^{\varepsilon'} \left(Z_{A}|ER\right)+2-2 \log_{2} \frac{1}{\varepsilon} \label{length}
\end{eqnarray}
can be extracted for any $\varepsilon_s, \varepsilon, \varepsilon' \geq 0$ such that 
\begin{eqnarray}
\varepsilon_s \geq \varepsilon+2\varepsilon' \label{epcond} 
\end{eqnarray}
where $H_{\min}^{\varepsilon'} \left(Z_{A}|ER\right)$ is the conditional smooth min-entropy of the raw measurement data given Eve's quantum system and the information reconciliation leakage conditioned on the protocol not aborting.
\end{corollary}

{\it Proof:} Note that if we choose \begin{eqnarray}
\ell=H_{\min}^{\varepsilon'}\left(Z_{A}|ER\right)+2-2 \log_{2} \frac{(1-\varepsilon_{\mathrm{rob}})}{\varepsilon}, 
\end{eqnarray}then the right hand side of (\ref{TD}) is equal to $\varepsilon/(1-\varepsilon_{\mathrm{rob}}) + 2 \varepsilon'$. Comparing with (\ref{es}) in Definition~\ref{secdef} we see we want this expression to satisfy $\varepsilon/(1-\varepsilon_{\mathrm{rob}}) + 2 \varepsilon' \leq \varepsilon_s/(1-\varepsilon_{\mathrm{rob}})$ so our security condition is satisfied for any $\varepsilon_s \geq \varepsilon+2(1-\varepsilon_{\mathrm{rob}})\varepsilon'$ which is true for any $\varepsilon_s \geq \varepsilon+2\varepsilon'$ where we used that $(1-\varepsilon_{\mathrm{rob}}) \leq 1$. Noting further that $\log_{2}(1-\varepsilon_{\mathrm{rob}}) \leq 0$ yields (\ref{length}). This means that, provided the constraint in (\ref{epcond}) is satisfied, the positive constant $\varepsilon$ can be optimised over. Typically this makes little difference to the final performance and and they are commonly chosen as $\varepsilon = \varepsilon_s/2$.

Now we see that the problem has condensed to determining Eve's conditional smooth min-entropy for $Z_A^k$ (in the following we will suppress the $k$ superscript), the variable describing the outcome of Alice's $\sigma_z$ measurements on the $k$ key-generating qubits. To begin with, consider the situation before any information reconciliation is exchanged (there is no register $R$) so we simply have $\hmin(Z_A|E)$. Since Eve's state is taken to include that of all the non-participating players we can assume without loss of generality that there is an overall pure tripartite state between Alice, the remaining participants (which we denote $B_i$), and Eve. The required bound for this situation has been derived by applying an entropic uncertainty relation \cite{Tomamichel:2011ci} for the smoothed min- and max-entropies specialised to the case of observables made up of the $k$-fold tensor product of either $\sigma_z$ and $\sigma_x$ measurements, (i.e. the observables $Z_A = \sigma_z^1\otimes\sigma_z^2\otimes\cdots\otimes\sigma_z^k$ and $X_A = \sigma_x^1\otimes\sigma_x^2\otimes\cdots\otimes\sigma_x^k$) \cite{Tomamichel:2012ffa}
\begin{eqnarray}
\hmin^\varepsilon(Z_A|E) + \hmax^\varepsilon(X_A|B_i) &\geq& k, \nonumber \\
\Rightarrow \hmin^\varepsilon(Z_A|E) &\geq& k - \hmax^\varepsilon(X_A|B_i), \label{eur}
\end{eqnarray}
where we have used the data processing inequality, $\hmax^\varepsilon(X_A|X_{B_i}) \geq \hmax^\varepsilon(X_A|X_{B_i})$, in the second line. Naively, this bound cannot be evaluated since it is counterfactual, i.e. the $k$ qubits are always measured in the $\sigma_z$-basis so we have no direct access to $\hmax^\varepsilon(X_A|X_{B_i})$, which is the conditional max-entropy of the participants given their Pauli measurements if Alice had instead measured in the $\sigma_x$ in these $k$ rounds. However, since the parameter estimation and key generation rounds were selected at random then it has been shown that Serfling's bound can be applied to statistically bound the $\sigma_x$ correlation that would have been observed in the $k$ key generation rounds based upon those that were actually observed in the parameter estimation rounds. This is expressed in the following result.

\begin{lemma}[Lemma 3 in Ref.~\cite{Tomamichel:2012ffa}]
Let $k$ be the number of key generation rounds, $m$ be the number of parameter estimation rounds, $d_0$ a threshold on the number of errors that can be observed during parameter estimation without the protocol aborting and $\varepsilon'>0$.
\end{lemma}
\begin{eqnarray}
\hmax^{\varepsilon'}(X_A|X_{B_i}) \leq kh_2(d_0 + \mu(\varepsilon'(1-\varepsilon_{\mathrm{rob}}))), \hspace{2mm} where \hspace{2mm} \mu(\varepsilon):=\sqrt{\frac{m+k}{m k} \frac{m+1}{m} \ln \frac{1}{\varepsilon}} .\label{mubound}
\end{eqnarray}
Putting all of these results together we can prove the following security statement.

\begin{theorem}[Security statement]
If the anonymous CKA protocol defined above proceeds without aborting an $(\max_{i\in \{B,C\}}\ell_{\mathrm{EC}}^i,\varepsilon_c)$ error correction protocol and a two-universal hashing are successfully applied then an $(\varepsilon_s + \varepsilon_c)$-secure key of length
\begin{align}
\begin{split}
\ell &= k\left [ 1 - h_2\left (Q_{\mathrm{tol}} + \mu\left (\frac{\varepsilon_\mathrm{s} - \varepsilon}{2} \right ) \right )\right ] +2 -2 \log_{2} \frac{1}{\varepsilon}-\ell_{\mathrm{EC}} - \ell_{\mathrm{hEC}} - L\cdot h_2(p) -3 \\
&= L \left[\left(1-p\right)\left[1 - h_{2}\left(Q_{\mathrm{tol}} + \mu\left(\frac{\varepsilon_{\mathrm{s}} - \varepsilon}{2}\right)\right) - h_{2}\left(Q_{z}\right)\right] - h_{2}(p)\right] + \log_{2}(\varepsilon^{2}\varepsilon_{\mathrm{c}}) - 2
\label{lfinal}
\end{split}
\end{align}
can be anonymously extracted.
\end{theorem}
{\it Proof:} At the conclusion of the protocol we can immediately apply Corollary~\ref{corol1} to the $k$ round classical-quantum state $\rho^k_{Z_A E} = \mathrm{tr}_{B_i} (\ket{\Psi_{A B_i E}}\bra{\Psi_{AB_{i}E}})$ to extract an $\varepsilon_s$-secret key of length
\begin{eqnarray}
\ell=H_{\min}^{\varepsilon'} \left(Z_{A}|ER\right)+2-2 \log_{2} \frac{1}{\varepsilon}
\end{eqnarray}
for positive constants satisfying 
\begin{eqnarray}
\varepsilon_s \geq \varepsilon + 2(1-\varepsilon_{\mathrm{rob}})\varepsilon'. \label{poscons}
\end{eqnarray}
Now, because all of the communication involved in error reconciliation is one-time padded to ensure anonymity we have that $H_{\min}^{\varepsilon'} \left(Z_{A}|E,R\right) = H_{\min}^{\varepsilon'} \left(Z_{A}|E\right)$ by definition. This gives
\begin{eqnarray}
\ell &=& H_{\min}^{\varepsilon'} \left(Z_{A}|E\right)+2-2 \log_{2} \frac{1}{\varepsilon} \nonumber \\
&\begin{array}{cc}
 (\ref{eur}) \\
 \geq 
\end{array}& k - H_{\max}^{\varepsilon'} \left(X_{A}|X_{B_i}\right)+2-2 \log_{2} \frac{1}{\varepsilon} \nonumber \\
&\begin{array}{cc}
 (\ref{poscons}) \\
 \geq 
\end{array}& k - H_{\max}^{(\varepsilon_s - \varepsilon)/2/(1-\varepsilon_{\mathrm{rob}})} \left(X_{A}|X_{B_i}\right)+2-2 \log_{2} \frac{1}{\varepsilon} \nonumber \\
&\begin{array}{cc}
 (\ref{mubound}) \\
 \geq 
\end{array}& k - k h_2\left (Q_{\mathrm{tol}} + \mu\left (\frac{\varepsilon_s - \varepsilon}{2} \right ) \right ) +2-2 \log_{2} \frac{1}{\varepsilon}, 
\end{eqnarray}
where in the third line we have also used that $\hmax^{\varepsilon_1}(X|Y)< \hmax^{\varepsilon_2}(X|Y)$ for $\varepsilon_1>\varepsilon_2$. This string is guaranteed to be $\varepsilon_s$-secret and, by Theorem~\ref{ecth}, if the error correction process did not abort then the string is also $\varepsilon_c$-correct. However, this is not a fair representation of the performance of the protocol, since we had to use up the reservoir of pre-shared key for the basis choices and for one-time padding the error reconciliation information. Thus, to get the length of useable key we need to calculate how much remains after we have replenished the pre-shared strings necessary for the next protocol implementation. Subtracting off the seed for basis choices, $Lh_2(p)$, and the length of the error correction information and verification, $\ell_{\mathrm{EC}}$ and $\ell_{\mathrm{hEC}}$ and the $3$ bits for the abort step, gives (\ref{lfinal}). 

\section{Anonymity in the protocol}\label{append:anonymity}
This section is concerned with the anonymity of the protocol. We first define anonymity according to the definition presented in \cite{grasselliSecureAnonymousConferencing2022} and adapt it to the setting of our protocol. Most importantly for our analysis, we need to show that all public communication -- the announced measurement results -- is independent of the choice of participants. We do this by showing that they are uniformly random and uncorrelated.

\subsection{Definition of anonymity}
We base our definition of anonymity on \cite{grasselliSecureAnonymousConferencing2022} (Eqs.~$B7$ and $B13$) and adapt it to our setting. The definition is given in terms of the relation to an ideal output state of the protocol.

Let $\sigma_{\net{}|abc}$ be the ideal output state on the entire network, conditioned on a particular choice of participants $\partic{} = \{\Alice, \Bob, \Charlie\}$. Let $G \subset \nonpart{}$ be a random subset of nodes that does not include the participants and define $\sigma_{G|abc} = \tr_{\net \setminus G} [ \sigma_{\net{}|abc} ]$. For all sets $G$ and every other choice of participants $\partic{}'=\{\node{a'},~\node{b'},~\node{c'}\}$, $\sigma_{\net{}|abc}$ should have the property:
\begin{equation}\label{eq:anonproperty}
    \sigma_{G|abc} = \sigma_{G|a'b'c'}.
\end{equation}

This property ensures that the reduced state of any set of non-participants is independent of the choice of participants.

We define a protocol with output state $\rho_{\net|abc}$ to be \textit{$\varepsilon_{\mathrm{an}}$-anonymous}, if for every choice of participants $\partic{} = \{\Alice,\Bob,\Charlie\}$, we have:
\begin{eqnarray}
    \left\|\rho_{\nonpart{}|abc} - \sigma_{\nonpart{}|abc}\right\| \leqslant \varepsilon_{\mathrm{an}},
\end{eqnarray}
where $\sigma_{\nonpart{}|abc}$ is any state that fulfills property \eqref{eq:anonproperty}.

\subsection{Proof of anonymity}

In the proposed protocol, the output state $\rho_{\nonpart{}|abc}$ has several registers. The only non-trivial registers that need to be addressed are the ones containing the classical communication of all the measurement outcomes $\{\spmo{i}\} \cup \{\gemo{i}\}$. The reason is that the reduced quantum state of any dishonest party is the maximally mixed state, which is independent of the choice of participants, and therefore trivially fulfills \eqref{eq:anonproperty}. Moreover, all other parties do not hold a quantum register by the end of the protocol. 

In the remainder of this section we will show that there are no correlations between any of the announced measurement outcomes $\{\spmo{i}\} \cup \{\gemo{i}\}$, \ie that the outcome distribution is indistinguishable from that of the uniformly drawn announcements of the nodes $\node{1},\node{a}$ and $\node{c}$ during Protocols~\ref{prot:statepreparation} and~\ref{prot:ghzextraction}. We can then conclude that we have complete anonymity, \ie our protocol is $\varepsilon_{\mathrm{an}}$-anonymous for $\varepsilon_{\mathrm{an}} = 0$.

Since the state of the network always remains separable between the tri-partition of the nodes to the left of (and including) Alice, the nodes to the right of (and including) Charlie, and the nodes between (and including) Alice and Charlie, it suffices to show that there are no correlations within the measurement announcements associated with these three separate groups.
We show this absence of correlations only for the left set, since the argument applies analogously to the other two sets.
We first show this in the case of an honest-but-curious non-participant, followed by the case where a non-participant may actively deviate from the protocol.

\subsubsection{Honest-but curious setting}
Consider the stabilizer of the network state after all $CZ$ operations have been performed in Step 2a of Protocol~\ref{prot:statepreparation}. It is generated by the following collection of operators:
\begin{equation}
\begin{aligned}\label{eq:generatorsofstate}
    \sigma_{x}^{\qubt{1}}\sigma_{z}^{\qubb{2}}, \\
    \sigma_{z}^{\qubt{1}}\sigma_{x}^{\qubb{2}}\sigma_{z}^{\qubt{2}}, \\
    \{\sigma_{z}^{\qubt{i}}\sigma_{z}^{\qubb{i+1}}\}_{i = 2}^{a-2}, \\
    \{\sigma_{z}^{\qubb{i}}\sigma_{x}^{\qubt{i}}\sigma_{x}^{\qubb{i+1}}\sigma_{z}^{\qubt{i+1}}\}_{i = 2}^{a-2}, \\
    \sigma_{z}^{\qubb{a-1}}\sigma_{x}^{\qubt{a-1}}\sigma_{x}^{\qubb{a}}, \\
    \sigma_{z}^{\qubt{a-1}}\sigma_{z}^{\qubb{a}}.
\end{aligned}
\end{equation}
The measurement operator of all measurement outcomes together depends on $\beta_{1}$ as
% \begin{align*}
%     \tag{$\beta_{1} = 0$} \sigma_{x}^{\qubt{1}}\sigma_{y}^{\qubb{2}}\sigma_{x}^{\qubt{2}}\sigma_{x}^{\qubb{3}} \sigma_{x}^{\qubt{3}} \sigma_{y}^{\qubb{4}} \sigma_{x}^{\qubt{4}}\sigma_{x}^{\qubb{5}} \sigma_{x}^{\qubt{5}} \sigma_{y}^{\qubb{6}} \dots \sigma_{x}^{\qubt{a-1}}, \\
%     \tag{$\beta_{1} = 1$}
%     \sigma_{y}^{\qubt{1}}\sigma_{x}^{\qubb{2}}\sigma_{x}^{\qubt{2}}\sigma_{y}^{\qubb{3}} \sigma_{x}^{\qubt{3}} \sigma_{x}^{\qubb{4}} \sigma_{x}^{\qubt{4}}\sigma_{y}^{\qubb{5}} \sigma_{x}^{\qubt{5}} \sigma_{x}^{\qubb{6}}\dots \sigma_{x}^{\qubt{a-1}},
% \end{align*}
\begin{equation}\label{eq:Moperator}
    M = \begin{cases}
    \sigma_{y}^{\qubb{2}}\sigma_{x}^{\qubt{2}}\sigma_{x}^{\qubb{3}} \sigma_{x}^{\qubt{3}} \sigma_{y}^{\qubb{4}} \sigma_{x}^{\qubt{4}}\sigma_{x}^{\qubb{5}} \sigma_{x}^{\qubt{5}} \sigma_{y}^{\qubb{6}} \dots \sigma_{x}^{\qubt{a-1}}, & (\beta_{1} = 0)\\
    \sigma_{x}^{\qubb{2}}\sigma_{x}^{\qubt{2}}\sigma_{y}^{\qubb{3}} \sigma_{x}^{\qubt{3}} \sigma_{x}^{\qubb{4}} \sigma_{x}^{\qubt{4}}\sigma_{y}^{\qubb{5}} \sigma_{x}^{\qubt{5}} \sigma_{x}^{\qubb{6}}\dots \sigma_{x}^{\qubt{a-1}}, &     (\beta_{1} = 1)
    \end{cases}
\end{equation}
\noindent
where all ($\sigma_{x}$-)observables acting on $\{\qubt{i}\}_{i=2}^{a-1}$ are associated with the measurements of Protocol~\ref{prot:statepreparation} (\ie the outcomes $\{\spmo{i}\}$) and all others are associated with Protocol~\ref{prot:ghzextraction} (\ie the outcomes $\{\gemo{i}\}$).

It is now our goal to show that all these measurement outcomes are uniformly random, and that there are no correlations between the measurement outcomes associated with any subset $S \subset Q$, where $Q = \{\qubb{2},\qubt{2}, \dots, \qubb{a-1},\qubt{a-1}\}$ 
is the set of qubits measured throughout both Protocol~\ref{prot:statepreparation} and Protocol~\ref{prot:ghzextraction}. Any such $S$ has an associated observable 
\begin{equation}
    M_{S} = \bigotimes_{i \in S} \sigma_{b(i)}^{i},
\end{equation}
where $b(i)\in\{x,y\}$ indicates the type of support on qubit $i$
%$\sigma_{b}^{i}$ is the support of $M$ on qubit $i$, \ie either $\sigma_{x}$ or $\sigma_{y}$ 
as shown in Eq.~\eqref{eq:Moperator}. If $M_{S}$ does not commute with 
at least one generator of the stabiliser (\ie any operator from Eq.~\eqref{eq:generatorsofstate}),  by Gottesman-Knill simulation, the measurement outcome for $M_{S}$ is uniformly random $0$ or $1$. If this holds for \textit{any} $S$, there cannot be any correlations between any of the measurement outcomes. The uniform randomness of the individual measurement outcomes follows readily for the case when $S$ contains only a single qubit. We now show that any $M_{S}$ indeed always anti-commutes with at least a single generator.

Suppose that $M_{S}$ \textit{does} commute with all generators but is non-trivial. If it has (non-trivial) support on $\qubt{a-1}$, this is necessarily with $\sigma_{x}$. It will then not commute with $\sigma_{z}^{\qubt{a-1}}\sigma_{z}^{\qubb{a}}$ (the last generator of Eq.~\eqref{eq:generatorsofstate}) and hence cannot have support on \qubt{a-1}. Then, if $M_{S}$ has (non-trivial) support on $\qubb{a-1}$, with either a $\sigma_{x}$ or $\sigma_{y}$, it will not commute with the generator $\sigma_{z}^{\qubb{a-1}}\sigma_{x}^{\qubt{a-1}}\sigma_{x}^{\qubb{a}}$ - thus it cannot have support on $\qubb{a-1}$ either.

We can inductively go through the rest of the qubits in $Q$ in reversed order, \ie from right to left through the observable from Eq.~\eqref{eq:Moperator}. For $j \in \{a-2,a-3, \dots, 3,2\}$:
\begin{itemize}
    \item[] Suppose $M_{S}$ has non-trivial support on \qubt{j}, it is of type $\sigma_{x}$. Since $M_{S}$ has by construction no support on any qubit to the right of \qubt{j}, it does not commute with the generator $\sigma_{z}^{\qubt{j}}\sigma_{z}^{\qubb{j+1}}$ - hence $M_{S}$ cannot have support on \qubt{j}.
    \item[] Suppose $M_{S}$ has non-trivial support on \qubb{j}, either of type $\sigma_{x}$ or $\sigma_{y}$. Since $M_{S}$ has by construction no support on any qubit to the right of \qubb{j}, it does not commute with the generator $\sigma_{z}^{\qubb{j}}\sigma_{x}^{\qubt{j}}\sigma_{x}^{\qubb{j+1}}\sigma_{z}^{\qubt{j+1}}$ -- hence $M_{S}$ cannot have support on \qubb{j}.
\end{itemize}

We conclude that there is no $M_{S}$ with non-trivial support on at least a single qubit that does not anti-commute with at least one generator.

From this, we can conclude that there are no correlations possible between any set of measurement outcomes from $\{\spmo{i}\}$ and $\{\gemo{i}\}$, and that they are thus uniformly random and uncorrelated. Moreover, it stays uniformly random under any noise that does not add a bias in the used measurement bases (\ie $\sigma_{x}$ and $\sigma_{y}$). 

\subsubsection{Dishonest participant}
We are now allowing a single non-participant to deviate from the protocol in an arbitrary way. Let the index of this dishonest non-participant be $i$. 
To try to force any other node in the network to implicitly reveal their identity, \node{i} can actively perform a different measurement than described, where their outcomes  would then be correlated with its (e.g.) direct neighbours. 
If these correlations then do not exist between their outcomes and the announced outcomes, then they can infer that these announced outcomes are artificial, and therefore that those who have announced them are in fact participants.
Let this arbitrary measurement be represented by a $2$-qubit POVM $\mu_{i} \defeq \{\mu_{i}^{j}\}$, where without loss of generality $j \in \{1,2,3,4\}$.

Slightly abusing notation by combining POVM elements and observables, the measurement operator then becomes

\begin{equation}\label{eq:Moperatordish}
    M = \begin{cases}
    \sigma_{y}^{\qubb{2}}\sigma_{x}^{\qubt{2}} \dots \sigma_{x}^{\qubb{i-1}} \sigma_{x}^{\qubt{i-1}} \bigotimes \mu_{i}^{j} \bigotimes \sigma_{x}^{\qubb{i+1}} \sigma_{x}^{\qubt{i+1}} \sigma_{y}^{\qubb{i+2}} \dots \sigma_{x}^{\qubt{a-1}}, & (\beta_{1} = 0)\\
    \sigma_{x}^{\qubb{2}}\sigma_{x}^{\qubt{2}} \dots \sigma_{y}^{\qubb{i-1}} \sigma_{x}^{\qubt{i-1}} \bigotimes \mu_{i}^{j} \bigotimes \sigma_{y}^{\qubb{i+1}} \sigma_{x}^{\qubt{i+1}} \sigma_{x}^{\qubb{i+2}} \dots \sigma_{x}^{\qubt{a-1}}, &     (\beta_{1} = 1).
    \end{cases}
\end{equation}

Without loss of generality, the underlying network state is still the same\footnote{Any non-trivial map that \node{i} may perform on their subsystem can be merged with the measurements $\{\mu_{i}\}$. The other participants don't deviate, or \node{i} is not aware of the deviation and therefore cannot exploit it.}
 as in \eqref{eq:generatorsofstate}. 
 Likewise, all of the single-qubit measurement operators in $M$ for any node \node{j \not = i} do not commute with at least one of these generators, indicating that the individual measurement outcomes are uniformly random $0$ or $1$.

Similar to before, the goal is to show that no choice of $\mu_{i}$ can create a measurement operator $M_{S}$ that shows correlations between the qubits of $i$ and any subset $S \subset Q$. It suffices to show that there is no $M_{S}$ with support on any of the qubits in $Q \setminus \{\qubt{i},\qubb{i}\}$ that commutes with all generators.
By the same analysis as in the previous section, $M_{S}$ cannot have any support on the qubits of any $\node{j|j \in  \{a-1, \dots, i+1\}}$. Moreover, we can make a similar inductive argument for nodes $\node{j|j \in \{2, \dots, i-1\}}$. Independent of $\beta_{1}$, if $M_{S}$ has support on $\qubb{2}$ it will not commute with the generator $\sigma_{x}^{\qubt{1}}\sigma_{z}^{\qubb{2}}$. 
Likewise, if $M_{S}$ has support on $\qubt{2}$, it will not commute with the generator $\sigma_{z}^{\qubt{1}}\sigma_{x}^{\qubb{2}}\sigma_{z}^{\qubt{2}}$. We can inductively go through all qubits from the nodes $\node{j|j \in \{2, \dots, i-1\}}$ to show that there exists no $M_{S}$ that has non-trivial support on any qubit of the nodes $\{2,\dots, i-1, i+1, \dots, a-1\}$ and at the same time commutes with all the generators. We can conclude that, even for a dishonest node \node{i}, there are no correlations in the measurement outcomes announced by the other nodes.

\newpage

\bibliography{resubmission_ACKA_Lin}

\end{document}